
\magnification=\magstep1          \def\sp{\,\,\,} \overfullrule=0pt
  \def\c{\chi} 
 \def\p{{\cal P}}
   \def\la{\lambda}   \def\R{{\cal K}} \def\Z{{\bf Z}}
\def\u{\tau}       \def\equi{\,{\buildrel \rm def \over =}\,}
\def\eg{{\it e.g.}$\sp$} \def\ie{{\it i.e.}$\sp$}
 \def\g{{\hat g}} \def\eps{\epsilon}
\def\d{\delta} \def\A{{\cal A}} \def\D{{\cal D}}
\def\E{{\cal E}}  \def\I{{\cal J}}
\font\huge=cmr10 scaled \magstep2
  \def\GW{[20]}  \def\COMM{[11]}  \def\BER{[2]}
 \def\CIZ{[6]} \def\RTW{[25]} \def\BOU{[3]} \def\STA{[27]}
\def\KAC{[22]} \def\MS{[24]} \def\GRS{[19]}  \def\Coit{[21,11]}
\def\HK{[14]}   \def\MAT{[17]} \def\Gko{[6,5,16]}
\def\SCH{[18]}  \def\GR{[13]} \def\Ht{[12,15]} \def\Cmtr{[14,15]}
\def\VT{[23]}  \def\SY{[26]} \def\BB{[1]} \def\BN{[4]} \def\Scpi{[18,10]}
\def\WN{[29]} \def\CL{[7]} \def\GWA{[16]} \def\CG{[8]} \def\Appl{[15,16]}
\def\FKSV{[10]}  \def\Df{[9,10]} \def\GH{[15]}
\def\Paru{[11,25,8]}\def\Wor{[2,4,26,28]}

{ \nopagenumbers
\rightline{February, 1994}
\bigskip \bigskip\bigskip\bigskip
\centerline{{\bf \huge Towards a Classification of su(2)$\bigoplus\cdots
\bigoplus$su(2)}}
\bigskip
\centerline{{\bf \huge Modular Invariant Partition Functions}}
\bigskip \bigskip\bigskip
\centerline{{Terry Gannon}}
\centerline{{\it Institut des Hautes Etudes Scientifiques,}}
\centerline{{\it 91440 Bures-sur-Yvette, France}}
\bigskip
\bigskip \bigskip \noindent{{\bf Abstract.}}
The complete classification of WZNW modular invariant partition functions is
known for
very few affine algebras and levels, the most significant being all levels of
$A_1$ and $A_2$ and level 1 of all simple algebras. Here, we address
the classification problem for the nicest high rank semi-simple affine
algebras:
$(A_1^{(1)})^{\oplus_r}$. Among other things, we explicitly find all
automorphism invariants, for all levels $k=(k_1,\ldots,k_r)$,
 and complete the classification for $A_1^{(1)}\oplus A_1^{(1)}$, for all
levels $k_1,k_2$. We also solve the classification problem for
$(A_1^{(1)})^{\oplus_r}$, for any levels $k_i$ with the property that
for $i\ne j$ each $gcd(k_i+2,k_j+2)\leq 3$. In addition, we find some
physical invariants which seem to be new. Together with some recent work
by Stanev, the classification for all $(A^{(1)}_1)^{\oplus_r}_k$ could
now be within sight.
\vfill \eject} \pageno=1

\noindent{{\bf 1. Introduction}} \bigskip

A great deal has been learned in recent years about rational conformal
field theories, but in spite of considerable effort relatively little
is known about their classification -- it is one of the main open
problems in the area. A natural first step is to try to classify all
Wess-Zumino-Novikov-Witten \WN{} models. The partition function of a
WZNW theory associated with algebra
 $\g$ and level $k$ can be
 written as a sesquilinear combination
$$Z=\sum M_{ab} \c_{ a}^{k}
\,\c_{b}^{k*} \eqno(1.1)$$
of characters $\c_{a}^{k}$  of the representation of
 $\g$ with (horizontal) highest weight $a$ and level $k$. Each
$\chi_{a}^k$ is a
function of a complex vector $z$ and a complex number $\tau$.
The algebra $\g$ is the untwisted
affine extension $g^{(1)}$ of a Lie algebra $g$ \KAC.

One curious reason the classifications of these $Z$ are interesting are the
mysterious ``coincidences'' which appear. For the $g=A_1$ classification,
there is the ADE pattern \CIZ; for the $g=A_2$ classification the Fermat
curves make surprise appearances \RTW. These are not fully understood at
present, and we can only guess as to their ultimate significance, but
they do encourage us to look a little deeper. It is in this spirit that
this paper has been written.

There are several conditions the function $Z$ in (1.1) must satisfy in order
to be realized by a physically reasonable conformal field theory. These are
discussed in Sect.2. When it does satisfy these conditions, we shall call
it a physical invariant.

Many tools (see \eg \Wor) have been developed over the past few years
for finding these physical
 invariants, and it is probably safe to say that almost all physical
invariants
are now known. But we have been far less successful at actually
proving that a given list exhausts all physical invariants belonging to
some choice of $g$ and $k$. The main {\it completeness proofs} at present
are: for $g=A_1$ and $g=A_2$, for any level $k$ (\CIZ{} and \GR, respectively);
and for $k=1$ for all simple $g$ \Coit. In addition,
some work in classifying {\it GKO} physical invariants
\Gko{} and {\it heterotic}
physical invariants \Ht{} has been done. The result for $A_1$ is:

\item{(1)} for each $k$, the diagonal invariant $\A_k$;

\item{(2)} for each even $k$, the complementary invariant $\D_k$;

\item{(3)} for $k=10,16,28$, the exceptionals $\E_{10}$, $\E_{16}$ and
$\E_{28}$.

\noindent{These} physical invariants can be found in Ref. \CIZ. ${\cal A}_k$
and ${\cal
D}_k$ can be thought of as simple current invariants \SY. ${\cal E}_{10}$
and ${\cal E}_{28}$ are due \BN{} to conformal embeddings, and ${\cal E}_{16}$
is due \MS{} to an exceptional automorphism of the ${\cal D}_{16}$ chiral
algebra. The $A_1$ result motivates much of the following. For example,
in Sects.3 and 4 we find the analogues of the ${\cal A}_k$, ${\cal D}_k$
invariants, and in Sect.5 we look for the analogues of ${\cal E}_{16}$.
In principle
all conformal embeddings are known \BB; they can occur for $A_{1,k_1}\oplus
\cdots\oplus A_{1,k_r}$ only when a $k_i=1,2,3,4,6,8,10$ or 28 (unfortunately,
{\it a priori} some exceptionals could be due to chiral extensions other than
conformal embeddings).

Unfortunately, the complexity increases with the rank, and when the algebra
is semi-simple the numbers of physical invariants, including exceptionals,
rises significantly. However, new methods developed in Ref. \GR, and
refined in this paper, make some of
the remaining classifications more accessible.
In this paper we will focus on the nicest algebra which is both
semi-simple and of arbitrarily high rank: namely,
$g^r\equi A_1\oplus\cdots\oplus A_1=(A_1)^{\oplus_r}$.
There are
several known physical invariants for it, which will be discussed in
Sect.2. The question this paper addresses is the completeness
of this list. This has also been addressed in other papers, \eg \Df.
All that has previously been established in this direction is that all
level $k=(k_1,k_2)$ physical invariants of $g^2$ have been found for small
$k_1,k_2$ by the computer search in Refs. \HK{} and \GH, and for odd $k_1$ and
$k_2$ by looking at operator subalgebras of ${\hat A}_{1,k}$ \CL{}.
Important recent progress \STA{} is the determination
of the possible chiral extensions of $g^r$, at least for small $r$.

In Sect.2 we describe most of the physical invariants of $g^r$,
summarize our main results, and list the physical invariants
of $g^2$. Sect.3 will find all
{\it permutation invariants} (see equation (3.1)) of $g^r$, for any
level $k=(k_1,\ldots,k_r)$. Some of these have not appeared in the literature
before. In Sect.4 we explicitly find all {\it simple current
invariants} (see equation (2.6)), and using this we find in Sect.5 all physical
invariants, for most $k$, due to automorphisms of simple current extensions.
We also find some new exceptionals (see equations (5.12)). In Sect.6
we illustrate some of this work by finding all physical invariants when the
heights $k_i+2$ are nearly coprime.
In Sect.7 we complete the classification for all levels of $g^2$.

There are a number of reasons why a classification of $A_1\oplus\cdots\oplus
A_1$ physical invariants would be interesting. For one thing, there is a
duality \GWA{} between the GKO cosets $G_k\oplus G_\ell/G_{k+\ell}$, and
a subset of the
WZNW $G_k\oplus G_\ell\oplus G_{k+\ell}$. In particular, knowing all
$g^3_{(k,\ell,k+\ell)}$ physical invariants will permit one to read off
all $A_{1,k}\oplus A_{1,\ell}/A_{1,k+\ell}$ physical invariants. For another
thing, it would be nice to know whether the intriguing ADE pattern \CIZ{}
for $A_1=g^1$ physical invariants continues in some way in $g^r$ for $r>1$.
So few complete classifications exist at present that a few more should give
us more hints of the global pattern applicable to all modular invariant
partition functions, which could ultimately lead to a more geometric global
understanding of conformal field theories, as well as of the various
connections between conformal field theories and other areas of mathematics
(\eg Fermat curves \RTW). $A_1\oplus\cdots\oplus A_1$ is more transparent
than any other large rank algebra, so understanding its physical invariants
should suggest new techniques or patterns to try on the others.
Finally, these $A_1\oplus\cdots\oplus A_1$ results are also of value in
classifying heterotic invariants \GH, and have been employed \FKSV{}
to produce Gepner-type compactifications of the heterotic string.

It should be mentioned that the rank-level duality $C_{n,k}\leftrightarrow
C_{k,n}$ \VT{} implies that our results for $\oplus_iA_{1,k_i}$ can be
carried over to $\oplus_iC_{k_i,1}$. This could have applications to the
search for exceptionals via conformal embeddings.

Although we obtain a complete classification only for $g^2$, the focus of
all but Section 7 is on arbitrary $g^r$. In particular, together with
\STA{}, the results of this paper should directly lead to classifications for
other $g^r$.

\bigskip \bigskip \noindent{{\bf 2. The physical invariants of $A_1\oplus
\cdots\oplus A_1$}}
\bigskip

We begin by introducing
notation and terminology. Our attention will be restricted here to the algebra
$g^r=(A_1)^{\oplus_r}$. Next, we will describe {\it most} of the physical
invariants of $g^r$. At the end of this section we summarize the main
results of this paper, and review the physical invariants of $g^2$.

Let
$\beta_1,\ldots,\beta_r$ denote the fundamental weights of $g^r$.
Throughout this paper we will identify the weight $a=a_1\beta_1+\cdots+
a_r\beta_r$ with
its Dynkin labels $(a_1,\ldots,a_r)$. Write $\rho_r=(1,\ldots,1)$.

An integrable irreducible  representation of the affine Lie algebra
$\g^r\equi (A_1^{(1)})^{\oplus_r}$ is associated with an $r$-tuple of positive
integers
$k=(k_1,\ldots,k_r)$ (called the {\it level}) and a {\it highest weight}
$a$ of $g^r$. For most purposes it will be more convenient to use the
{\it height} $k'=(k_1',\ldots,k_r')\equi(k_1+2,\ldots,k_r+2)$.
By $g^r_k$ we mean $g^r$ at level $k$.
The set of all possible highest weights corresponding to  level $k$
representations is
$$P_k^r\equi  \{(a_1,\ldots,a_r)\,|\,a_i\in\Z,\sp
0< a_i< k_i',\sp \forall i\}.\eqno(2.1)$$
The character corresponding to the level $k$ representation with
weight $a=(a_1,\ldots,a_r)\in P^{r}_k$ can be written
$$\chi^k_{a}={\tilde\chi}^{k_1}_{a_1}\cdots
{\tilde\chi}^{k_r}_{a_r},$$
where the $\tilde\chi$ denote $A_1$ characters. By including their full
variable dependence, for fixed $k$ the $\c_a^k$ will all be linearly
independent \KAC.

The function $Z$ in (1.1) can and will be identified with its coefficient
matrix $M$.
Three properties $Z$ must satisfy in order
to be the partition function of a physically sensible conformal field theory
are:
\medskip
\item{(P1)} {\it modular invariance}. Letting $S^{(k)}$ and $T^{(k)}$
denote the (unitary) modular matrices of $g^r_k$ (see equations (2.3) below),
this is equivalent to
the two conditions:
$$\eqalignno{T^{(k)\dag}MT^{(k)}=M, &\qquad i.e.\quad MT^{(k)}=T^{(k)}M,&
(2.2a)\cr
S^{(k)\dag}MS^{(k)}=M,&\qquad i.e. MS^{(k)}=S^{(k)}M; &(2.2b)\cr}$$

\item{(P2)} {\it positivity} and {\it integrality}. The coefficients
$M_{ ab}$ must be non-negative integers; and

\item{(P3)} {\it uniqueness of vacuum}.  We must have $M_{\rho_r\rho_r}=1.$
\medskip

We will call any modular invariant function $Z$ of the form (1.1),
an {\it invariant}. Together they define a complex space, called the
{\it commutant} $\Omega_k^r$. It is closed under matrix multiplication.
$Z$ will be called {\it positive} if
in addition each $M_{ab}\ge 0$, and {\it physical}
 if it satisfies (P1), (P2), and (P3).
Our task is to find all physical invariants corresponding
to each algebra $g^r$ and level $k$.

An invariant satisfying (P1), (P2) and (P3) is still not necessarily the
partition
function of a conformal field theory obeying duality. If it
is, we will call it {\it strongly physical}. These are the invariants of
interest to physics. We will discuss the additional properties
satisfied by strongly physical invariants (most importantly, that they
become automorphism invariants when written in terms of the characters of
their maximal chiral algebras) at the beginning of Sect.5.

In any classification it is desirable to use as few assumptions as possible.
For one thing this greater generality allows a greater opportunity for the
given classification to be directly relevant to other classifications,
some perhaps lying in entirely different domains (\eg ADE \CIZ, fermat curves
 \RTW).
In addition, there is a simple relationship between GKO coset physical
invariants and WZNW ones, and additional properties can disturb this
relationship. We will try as much as we can to restrict ourselves to
(P1)-(P3); when we need to exploit other properties we will clearly say.

The $\hat A_1$ characters ${\tilde\c}^k_a$ behave quite nicely under the
modular transformations $\u\rightarrow \u+1$ and $\u\rightarrow -1/\u$:
$$\eqalignno{{\tilde\c}_a^k(z,\u+1)=&\sum_{b\in P^1_k}\big({\tilde{T}}^{(k)}
\big)_{ab}{\tilde\c}_{b}^{k}(z,\u),\sp \sp{\rm where}&(2.3a)\cr
\big({\tilde{T}}^{(k)}\big)_{ab}=&\exp[\pi i{ a^2\over 2(k+2)}-\pi i
{1\over 4}] \,\delta_{ab}\sp; &(2.3b)\cr
\tilde{\c}_a^{k}(z/\u,-1/\u)=&\exp[k\pi iz^2/\u]\,\sum_{b\in P^1_k}\big(
{\tilde{S}}^{(k)}\big)_{ab}{\tilde{\c}}_{b}^{k}(z,\u),\sp \sp{\rm where}&
(2.3c)\cr\big({\tilde{S}}^{(k)}\big)_{ab}=&\sqrt{{2\over k+2}}
\sin[\pi  {ab\over k+2}]. &(2.3d)\cr}$$
For arbitrary rank $r$, the corresponding modular matrices $T^{(k)}$ and
$S^{(k)}$ become $\tilde{T}^{(k_1)}\otimes\cdots\otimes \tilde{T}^{(k_r)}$ and
$\tilde{S}^{(k_1)}\otimes \cdots \otimes \tilde{S}^{(k_r)}$ respectively; for
all $r$ and $k$ they are real, orthogonal and symmetric.

The easiest source of invariants for $g^r_k$ is through the matrix tensor
product $M\otimes M'$. In particular, let $k^{1s}=(k_1,\ldots,k_s)$ and
$k^{sr}=(k_{s+1},\ldots,k_r)$. If $M$ is a physical invariant of $g^s$
level $k^{1s}$
and $M'$ is a physical invariant of $g^{r-s}$ level $k^{sr}$, then $M\otimes
M'$ will be a physical invariant of $g^r$ level $k$. But in general
this fails to construct most physical invariants of $g^r_k$.

Another important source of physical invariants is given by simple
currents \SY, \ie outer automorphisms \BER{} of $g^r_k$.
For $g^r_k$, a simple current $J$ is just an $r$-tuple of 0's and 1's. They
form a group, denoted $\I$, under component-wise addition (mod 2), hence a
vector space
over ${\bf Z}_2$, the integers mod 2. $J$ acts on a weight $a\in P^r_k$ by
$$(Ja)_i=\left\{\matrix{a_i&{\rm if}\sp J_i=0\cr k_i'-a_i&{\rm if}\sp
J_i=1\cr}\right. .\eqno(2.4a)$$
Simple currents play a central role in this paper. Their
norms, defined mod 4, and inner products, defined mod 2, are given by
$$\eqalignno{J^2&\equi\sum_{i=1}^r (J_i)^2k_i\qquad ({\rm mod}\sp 4),&(2.4b)\cr
J\cdot J'&\equi\sum^r_{i=1} J_iJ_i'k_i\qquad ({\rm mod}\sp 2).&(2.4c)\cr}$$
Another useful quantity is $J\cdot (a-\rho_r)=\sum_i J_i(a_i-1)$, defined
mod 2.

Each simple current $J$ with even norm $J^2$ defines an {\it elementary
simple current invariant} $M(J)$: for $J^2\equiv 0$ (called integer-spin),
it is given by
$$M(J)_{ab}=\left\{\matrix{\delta_{ab}+\delta_{a,Jb}&{\rm if}\sp J\cdot
(a-\rho_r)\equiv 0\sp ({\rm mod}\sp 2)\cr 0&{\rm otherwise}\cr}\right. ;
\eqno(2.5a)$$
and for $J^2\equiv 2$ (called half-integer spin), by
$$M(J)_{ab}=\left\{\matrix{\delta_{ab}&{\rm if}\sp J\cdot(a-\rho_r)\equiv 0\sp
({\rm mod}\sp 2)\cr \delta_{b,Ja}&{\rm otherwise}\cr}\right. .\eqno(2.5b)$$
We will usually write the invariant $M(J)$ in $(2.5b)$ as $I^J$, for
reasons which will be clearer next section.

More generally, call a physical invariant $M$ a {\it simple current
invariant} \SCH{} if for all $a,b$ it obeys the selection rule:
$$M_{ab}\ne 0 \Rightarrow b=Ja,\eqno(2.6)$$
for some simple current $J$ (depending on $a,b$).
Elementary simple current invariants are only a small subset of these,
but we will find in Sect.4 that they need only a little help to
generate all simple current invariants.

When some of the levels $k_i$  are equal, we may obtain
new physical invariants  from old ones through {\it conjugation}.
Call a permutation $\pi$ of $\{1,\ldots,r\}$ a conjugation if $k_i=
k_{\pi i}$ for all $i=1,\ldots,r$. Then for any physical invariant $M$,
we define its conjugation $M^\pi$ by the formula
$$\bigl(M^\pi\bigr)_{a,b}=M_{\pi a,b}.\eqno(2.7)$$
For $r=2$ the notation $M^c$ is common. This conjugation is not to be mistaken
for {\it charge conjugation}: the charge conjugation matrix $C=S^{(k)2}$ for
 $g^r_k$ is just the identity.

Together, simple current invariants and their conjugations constitute what may
be called the {\it regular series} of physical invariants, and can be expected
to exhaust almost all physical invariants of $g^r$. For a given $g^r_k$ their
number is given by the following remarkable formula \GRS
$$\big(\prod_{\ell>1} n_\ell!\big)\,\prod_{i=0}^{r'} (2^i+1),\eqno(2.8)$$
where $n_\ell$ is the number of $k_i=\ell$, and $r'=r-1$ or $r-2$,
depending on whether or not all $k_i$ are even (if some $k_i=2$, there
will be some overcounting in (2.8)).

By an {\it exceptional invariant} we mean any physical invariant which is
not ``regular''. For $r>1$, infinite series of exceptionals exist. They
can be constructed for example by taking an exceptional for $g^s$, where $s<r$,
and tensoring it with any physical invariant from $g^{r-s}$. But there
may be some exceptionals which cannot be built up from lower rank
exceptionals -- we will call these {\it sporatic exceptionals}. $g^1$
has 3 sporatics, and $g^2$ has 8 (12, if one treats \eg $k=(2,10)$
as different from (10,2)). The possible existence of sporatics
complicates any attempt at classification.

Our main results are (more careful statements of them, along with all
relevant definitions, can be found later in the paper):

\medskip
\noindent{{\bf Theorem 1.}}\quad{\it Any permutation invariant of any $g^r_k$
is generated, through matrix and tensor products, by conjugations $\pi$,
by elementary simple currents $I^J$, and by three
different families of ``integer-spin'' simple current automorphisms.}

\noindent{{\bf Theorem 2.}}\quad{\it Any simple current invariant can be
written as
the product of a permutation invariant with a number of elementary simple
current invariants $M(J)$.}

\noindent{{\bf Theorem 4.}}\quad {\it Provided five anomolous levels are
avoided,
only simple current invariants and their conjugations obey the following
property:  the only weights $a$ coupled to $\rho_r$ (\ie satisfying
$M_{a\rho_r}\ne 0$ or
$M_{\rho_r a}\ne 0$) are of the form $a\in\I\rho_r$}.

\noindent{{\bf Theorem 7.}}\quad {\it Provided each $gcd(k_i',k_j')\leq 3$
for $i\ne j$, then all physical invariants of $g^r_k$ are known.}

\noindent{{\bf Theorem 8.}} \quad {\it All physical invariants of $g^2_k$, for
any $k=(k_1,k_2)$, are known.}
\medskip

Here is a summary of all the $g^2_k$ physical invariants. Let ${\cal A}_k$,
${\cal D}_k$ and ${\cal E}_k$ denote the physical invariants of $A_{1,k}$.
There are either 6 or 2 simple current series for $g^2_k$, depending on
whether or not both $k_1,k_2$ are even. These are explicitly given in the
Appendix of
\GH.  When $k_1$ equals 10, 16 or 28, additional physical invariants
are the tensor product ${\cal E}_{k_1}\otimes {\cal A}_{k_2}$ for all $k_2$,
and the tensor product ${\cal E}_{k_1}\otimes {\cal D}_{k_2}$ and the matrix
product $({\cal E}_{10}\otimes{\cal A}_{k_2})\, M(J)$ for $J=(1,1)$
if $k_2$ is even. Similarly if instead $k_2=10$, 16 or 28.
If both $k_1,k_2\in\{10,16,28\}$, then in addition there is ${\cal E}_{k_1}
\otimes {\cal E}_{k_2}$. If $k_1=k_2$, the conjugations of all those
invariants must also be included,
of course. The sporatic exceptionals for $A_1\oplus A_1$ are given in Refs.
\Cmtr: there is exactly one at each level
$(k_1,k_2)$=(4,4), (6,6), (8,8), (10,10), (2,10), (3,8), (3,28) and (8,28).
In particular, see equations $(4.3f)$-$(4.3i$) of \HK{}\footnote*{The
(4,4)
exceptional appeared incorrectly in an early preprint of \HK; the correct
invariant reads: $|\c_{11}+\c_{15}+\c_{51}+\c_{55}|^2+
|\c_{13}+\c_{31}+\c_{35}+\c_{53}|^2+4|\c_{33}|^2$.} and
$(A.6)$-$(A.9)$ of \GH.

At the end of Sect.7 we list the numbers of physical invariants of $g^2$,
for each $k$.

\bigskip \bigskip \noindent{{\bf 3. The permutation invariants}}
\bigskip

By a {\it permutation invariant} (or {\it automorphism
invariant}) we mean a physical invariant of the form
$$\eqalignno{Z=&\sum_{a\in P^r_k} \c_a \, \c_{\sigma a}^*, &(3.1a)\cr
i.e.\sp M_{ab}=&\bigl(I^\sigma\bigr)_{ab}\equi
\delta_{b,\sigma a}&(3.1b)\cr}$$
for some permutation $\sigma$ of $P^r_k$. (P3) says
$\sigma(\rho_r)=\rho_r$. In this section we will find all
$g^r_k$ permutation invariants, for each rank $r$ and level $k$.

Recall that we write $k'=(k_1',\ldots,k_r')=(k_1+2,\ldots,k_r+2)$.
For convenience we will often drop the subscript on $\rho_r$.

First let us list some simple examples of $g^r_k$ permutation invariants.
The trivial one corresponds to the identity matrix $M=I$. Other examples
are its conjugations $M=I^\pi$ (see (2.7)).

Another source of permutation invariants are the $J^2\equiv 2$ (mod 4) simple
currents (see $(2.5b)$).  Matrix products of these $I^J$ produce many other
simple current invariants (2.6) which are permutation invariants. But they
do not generate all of them:
there are also {\it integer-spin simple current permutation
invariants} \Scpi{}. For example, choose any $m\ne n$ with $k_m\equiv
k_n\equiv 0$ (mod 4), and define a matrix $I^{mn}$ by
$$\bigl(I^{mn}\bigr)_{ab}=\bigl(\prod_{i\ne m,n} \delta_{a_ib_i}\bigr)\cdot
\left\{
\matrix{\delta_{a_mb_m}\delta_{a_nb_n}&{\rm if}\sp a_m\equiv a_n\equiv 1
\sp {\rm (mod}\sp 2)\cr
\delta_{k_m'-a_m,b_m}\delta_{a_nb_n}&{\rm if}\sp a_m\equiv 1,\sp a_n\equiv 0
\sp {\rm (mod}\sp 2)\cr
\delta_{a_mb_m}\delta_{k_n'-a_n,b_n}&{\rm if}\sp a_m\equiv 0,\sp a_n\equiv 1
\sp {\rm (mod}\sp 2)\cr
\delta_{k_m'-a_m,b_m}\delta_{k_n'-a_n,b_n}&{\rm if}\sp a_m\equiv a_n\equiv 0
\sp {\rm (mod}\sp 2)\cr}\right. .
\eqno(3.2)$$
Then $I^{mn}$ will be a permutation invariant. It was first found in \Scpi.

There are two further examples. One,
for $r>2$, involves a tensor product of a $g^3$
permutation invariant with the identity. Choose any $\ell,m,n$ with $k_\ell
\equiv 1$, $k_m\equiv -1$, and $k_n\equiv 0$ (mod 4), and define a matrix
$$\eqalignno{\bigl(I^{\ell mn}\bigr)_{ab}=\bigl( \prod_{i\ne \ell,m,n}
\delta_{a_ib_i} \bigr)\cdot
&\left\{\matrix{\delta_{a_\ell b_\ell}\delta_{a_mb_m}&{\rm if}\sp a_n\equiv
1\sp({\rm mod}\sp 2)\cr \delta_{k_\ell'-a_\ell,b_\ell}\delta_{k_m'-a_m,b_m}
&{\rm if}\sp a_n\equiv 0\sp({\rm mod}\sp 2)\cr}\right. &\cr
&\times \left\{
\matrix{\delta_{a_nb_n}&{\rm if}\sp a_\ell\equiv a_m
\sp({\rm mod}\sp 2)\cr \delta_{k_n'-a_n,b_n}
&{\rm if}\sp a_\ell\not\equiv a_m\sp({\rm mod}\sp 2)\cr}\right. .&(3.3)\cr}$$
The other, for $r>3$, involves a tensor product of a $g^4$ permutation
invariant with the identity. Choose any $\ell,m,n,p$ with $k_\ell\equiv k_m
\equiv -k_n\equiv -k_p\equiv 1$ (mod 4), and define a matrix
$$\eqalignno{\bigl(I^{\ell mnp}\bigr)_{ab}=\bigl( \prod_{i\ne \ell,m,n,p}
\delta_{a_ib_i}\bigr)\cdot
&\left\{\matrix{\delta_{a_\ell b_\ell}\delta_{a_pb_p}&{\rm if}\sp a_m\equiv
a_n\sp({\rm mod}\sp 2)\cr \delta_{k_\ell'-a_\ell,b_\ell}\delta_{k_p'-a_p,b_p}
&{\rm if}\sp a_m\not\equiv a_n\sp({\rm mod}\sp 2)\cr}\right. &\cr
&\times \left\{
\matrix{\delta_{a_mb_m}\delta_{a_nb_n}&{\rm if}\sp a_\ell\equiv a_p
\sp({\rm mod}\sp 2)\cr \delta_{k'_m-a_m,b_m}\delta_{k_n'-a_n,b_n}
&{\rm if}\sp a_\ell\not\equiv a_p\sp({\rm mod}\sp 2)\cr}\right. .&(3.4)\cr}$$
Both $I^{\ell mn}$ and $I^{\ell mnp}$ are also simple current permutation
invariants.
They
cannot be expressed as (tensor or matrix) products of the permutation
invariants $I^\pi$, $I^J$ and $I^{mn}$ listed earlier, or of each other,
and seem to have never appeared explicitly in the literature before.

\medskip\noindent{{\bf Theorem 1.}} \quad {\it Any permutation
invariant $I^\sigma$ of $g^r_k$ can be written as the matrix product}
$$I^\sigma=\bigl(\prod_{J\in {\cal A}} I^J\bigr)\,\bigl(\prod_{(m,n)\in {\cal
B}}I^{mn}\bigr)\,\bigl(\prod_{(\ell,m,n)\in {\cal C}}I^{\ell mn}\bigr)\,
\bigl(\prod_{(\ell,m,n,p)\in {\cal D}}I^{\ell mnp}\bigr)\,I^\pi.\eqno(3.5)$$

Any or all of
the sets ${\cal A},{\cal B},{\cal C},{\cal D}$ may be empty, in which case the
corresponding product is defined to be the identity matrix $I$. Of course,
${\cal A}$
is a set of simple currents $J$ with $J^2\equiv 2$, ${\cal B}$ is a set of
pairs
$(m,n)$ with $m\ne n$ and $k_m\equiv k_n\equiv 0$ (mod 4), ${\cal C}$ is a
set of triplets $(\ell,m,n)$ with $k_\ell\equiv 1$, $k_m\equiv -1$, and
$k_n\equiv 0$ (mod 4), ${\cal D}$ is a 4-tuple $(\ell, m,n,p)$ satisfying
$\ell\neq m$, $n\neq p$, $\ell\equiv m\equiv -n\equiv-p\equiv 1$ (mod 4),
and $\pi$ is a conjugation.

The set of all permutation invariants of $g^r_k$ forms a (generally nonabelian)
group.  We will prove below that the $I^J$, $I^{mn}$, $I^{\ell mn}$, $I^{\ell
mnp}$ and $I^\pi$ generate this group. From this, it is easy to show that
any $I^\sigma$ can be written as in (3.5).

An alternate description of the permutation invariants of $g^r_k$ is
given in Lemma 1 proved below. The remainder of this section is devoted
to a proof of Thm.1. The proof will be quite explicit, since it will be
exploited later in the paper.

Throughout this section we will write $a'$ for $\sigma a$, $b'$ for $\sigma b$,
etc.

That the matrix $I^\sigma$ in (3.1$b$) must commute with $S^{(k)}$
and $T^{(k)}$ (see (P1)) is equivalent to
$$\eqalignno{\prod_{i=1}^r \sin(\pi a_ib_i/k'_i)&=\prod_{i=1}^r \sin(\pi a_i'
b_i'/k_i'),&(3.6a)\cr
\sum_{i=1}^r {a_i^2\over k'_i}&\equiv\sum_{i=1}^r {a_i'{}^2\over k_i'}\sp
({\rm mod}\sp 4),&(3.6b)\cr}$$
for all $a,b\in P_k^r$.

The fusion coefficients $\tilde{N}^{(k)}_{lmn}$ of $A_1$ level $k$ are
well-known \GW:
$$\tilde{N}^{(k)}_{lmn}=\left\{ \matrix{1&{\rm if}\sp l+m+n\equiv 1\sp(
{\rm mod\sp 2)\sp and\sp} |l-m|<n<\min\{l+m,2k'-l-m\}     \cr
0&{\rm otherwise}\cr}\right. .\eqno(3.7a)$$
Of course, the fusion coefficient $N^{(k)}_{abc}$ of $g^r_k$ is just the
product
$$N^{(k)}_{abc}=\prod_{i=1}^r \tilde{N}^{(k_i)}_{a_ib_ic_i}.\eqno(3.7b)$$

Verlinde's formula implies the useful fact $N^{(k)}_{abc}=N^{(k)}_{a'b'c'}$.
Let $f_k(m)$ be the number of $n$ such that $\tilde{N}^{(k)}_{mmn}=1$, then
from (3.7$a$) we find that $f_k(m)=$min$\{m,k+2-m\}$. But from
(3.7$b$), for any $a\in P^r_k$ the number of $b\in P_k^r$ such that
$N^{(k)}_{aab}=1$ equals $f_{k_1}(a_1)\cdots f_{k_r}(a_r)$. Thus we get
the important equality:
$$\prod_{i=1}^r {\rm min}\{a_i,k_i'-a_i\}=\prod_{j=1}^r {\rm min}\{a_j',
k_j'-a_j'\}.\eqno(3.8)$$

The first step in the proof of Thm.1 is to ``factor off'' the
conjugation $I^\pi$ in (3.5).

\medskip\noindent{\bf Claim 1.} \quad {\it Let $e^i=(1,\ldots,2,\ldots,1)\in
P^r_k$ be the weight whose $j$th component is $(e^i)_j=1+\delta_{ij}$. Let
$e^i{}'=\sigma e^i$. Then $(e^i{}')_j=J e^{\bar{i}}$, for some simple current
$J$ (depending on $i$), and some index $\bar{i}$ satisfying $k_{\bar{i}}=k_i$.}

This is trivial for $k_i=1$; for future convenience choose $\bar{i}=i$
when $k_i=1$.

\medskip\noindent{\it Proof.} \quad We may assume $k_i>1$.
Putting $a=e^i$ in (3.8), we see that
the LHS equals 2, which forces exactly one of the factors on the RHS (call it
$\bar{i}$) to equal 2, and the remaining factors to equal 1. In other
words, for each $j$ either $(e^i{}')_j=1+\d_{j,\bar{i}}$ or $k'_j-(1+
\d_{\bar{i},j})$.

It suffices to show $k_{\bar{i}}=k_i$. This follows from $(3.6a$) with
$a=e^i$
and $b=\rho$: we get $\sin(\pi/k'_{\bar{i}})\cdot\sin(2\pi/k_i')=\pm
\sin(\pi/k_i')\cdot\sin(2\pi/k'_{\bar{i}})$, \ie $\cos(\pi/k_i')=\pm
\cos(\pi/k'_{\bar{i}})$ and hence $k'_i=k'_{\bar{i}}$.
\qquad QED to claim \medskip

Claim 1 defines a function $\pi$ of $\{1,\ldots,r\}$, by $\pi i=\bar{i}$.
We would like to show $\pi$ is a {\it conjugation} of $g^r_k$. To do
this it suffices now to show that $\pi$ is a permutation -- \ie if $\pi i=\pi
j$, then $i=j$.

Suppose for some $i\ne j$, $\pi i=\pi j$. Then by Claim 1,
$k_i=k_j=k_{\pi i}=k_{\pi j}$.
{}From $(3.6a$) with $a=e^i$ and $b=e^j$ we get $\sin(2\pi/k_i')\cdot
\sin(2\pi/k_i')=\pm \sin(4\pi/k_i')\cdot \sin(\pi/k_i')$, \ie $\cos(\pi/k_i')
=\pm \cos(2\pi/k_i')$, which only has one solution: $k_i'=3$. But we have
defined $\pi i=i$ whenever $k_i=1$, so $i=\pi i=\pi j=j$.

Therefore $\pi$ is a conjugation. Multiplying $I^\sigma$ on the right
by the conjugation $I^{\pi^{-1}}$ yields a
permutation invariant $I^{\sigma'}$ whose $\pi$ is the identity.
Thus we may assume without loss of generality in the following that $\sigma$
is such that $\bar{i}=i$ in Claim 1. This will make our
notation somewhat cleaner.

Comparing $b=e^i$ with $b=\rho$
in $(3.6a$), we see that for any $a$, $a'_i$ either equals
$a_i$ or $k'_i-a_i$. Hence $I^\sigma$ is a simple current invariant (see
(2.6)). The simple current permutation invariants
 have been completely classified in Ref.~\SCH{} (subject however to an
assumption
on the $S$-matrix which does not appear to be satisfied by most $g^r_k$). Our
argument, from here to Lemma 1, resembles that of \SCH, but
because we restrict attention here to $g^r$ we are able to explicitly solve
the equations of Lemma 1, and obtain in the end equation (3.5).

We have shown that a permutation invariant is completely specified by
the simple currents $J^a$ defined for each $a$ by $a'=J^aa$.  Define $f^i=
J^{e^i}$ and ${\tilde f}^i\in \I$ by $\sigma^{-1}(e^i)=\tilde{f}^i e^i$.
This definition of $J^a_i$ is
ambiguous when $k'_i=2a_i$ --- we are free then to choose any value for
$J^a_i$. It is most convenient however to fix them using equation (3.9) below.
When $k_i=2$, $f^i_i$ and $\tilde{f}^i_i$ may be arbitrarily chosen.

For any $a\in P_k^r$ and each $i=1,\ldots,r$, putting $b=\sigma^{-1}(e^i)$ in
$(3.6a$) gives us
$$J^a_i\equiv (a-\rho)\cdot {\tilde f}^i \equi \sum_{j=1}^r (a_j-1)
{\tilde f}^i_j\qquad ({\rm mod}\sp 2).\eqno(3.9)$$
Indeed, we get $\pm \prod_{j=1}^r\sin(\pi(e^i)_ja_j/k_j')=\pm
\prod_{j=1}^r\sin(\pi(e^i)_ja_j/k_j')$, where the LHS sign is
$(-1)^{(a-\rho)\cdot{\tilde f}^i}$ and the RHS sign is $(-1)^{J^a_i}$. The
product of $\sin$'s vanishes iff $2a_i=k'_i$, in which case
as we noted above we may freely choose (3.9) to hold.

Hence $f_i^j={\tilde f}_j^i$, which says that the $f$-matrix for $\sigma^{-1}$
is the transpose of the $f$-matrix for $\sigma$. Also, we have learned that
 $\sigma$ is completely specified by the $f^i$.

Rewriting equation ($3.6b$) gives us $\sum_i k'_i(J^a_i)^2\equiv 2 a\cdot J^a$
(mod 4). Inserting $a=e^i$ gives us
$$\sum_{j=1}^r k_j'f^i_j\equiv 2\sum_{j=1}^r f^i_j+2f^i_i\qquad({\rm mod}\sp 4)
.\eqno(3.10a)$$
For $i\ne j$, inserting next $a=e^i+e^j-\rho$ and using (3.9) and $(3.10a$)
gives us
$$f^i_j+f^j_i\equiv \sum_\ell k_\ell f^i_\ell f^j_\ell \qquad ({\rm mod}\sp
2).\eqno(3.10b)$$
Equations (3.9),(3.10) suffice to prove (3.6$b$) for
arbitrary $a$, so we have extracted all the information we can from it.
Similarly, (3.9) and $(3.10b$) suffice to prove $(3.6a$). Summarizing
our results:

\medskip\noindent{\bf Lemma 1.}\quad {\it To find all possible permutation
invariants
$I^\sigma$ of $g^r_k$, first find all possible conjugations $\pi$; then
find all possible $r\times r$ matrices $F=\bigl(f_{ij}\bigr)$ of 0's and 1's
such that:}

\item{(i)} $f_{ij}+f_{ji}\equiv \sum_\ell k_\ell f_{i\ell}f_{j\ell}\sp$ (mod
2), $\forall i,j$;

\item{(ii)} $\sum_j k_j f_{ij}\equiv 2f_{ii}\sp$ (mod 4), $\forall i$.

{\it To each such $F$ define a permutation $\sigma_F$ by}
$$\sigma_F(a)_i=\left\{ \matrix{a_i&{\rm if}\sp \sum_{j=1}^r (a_j-1)f_{ji}
\equiv 0\sp({\rm mod}\sp 2)\cr k_i'-a_i&{\rm otherwise}\cr}.\right.
\eqno(3.11)$$
{\it Then to each pair $(F,\pi)$ there is a permutation invariant given by
$I^{\sigma_F}I^\pi$. This exhausts all permutation invariants of $g^r_k$.
Provided all $k_i\ne 2$ and at most one $k_i=1$, then to each different
$(F,\pi)$ will correspond a different permutation invariant.}\medskip

Actually, we have not yet shown that $\sigma_F$ will be a permutation of
$P^r_k$. But if $J^aa=J^bb$, then using (3.9) and $(3.10b$) to expand out
$J^{e^\ell}\cdot(J^aa-\rho)\equiv J^{e^\ell}\cdot (J^bb-\rho)$, we get
$J^a_\ell=J^b_\ell$, $\forall \ell$. Hence $J^aa=J^bb$ implies $J^a=J^b$,
and thus $a=b$. So $\sigma_F$ will indeed be a permutation, and define
a permutation invariant.

But whenever $\sigma$ is a permutation invariant, so will be $\sigma^{-1}$.
We know that the $F$-matrix for $\sigma^{-1}$ is the transpose of that for
$\sigma$. From this we get two other formulas any $F$ satisfying (i) and (ii)
must satisfy:
$$\eqalignno{f_{ij}+f_{ji}&\equiv \sum_{\ell=1}^r k_\ell f_{\ell i}
f_{\ell j}\qquad {\rm (mod}\sp 2)\qquad \forall i,j;&(3.12a)\cr
\sum_{j=1}^r k_j f_{ji}&\equiv 2f_{ii}\qquad ({\rm mod}\sp 4)\qquad \forall
i.&(3.12b)\cr}$$

It remains to express each $\sigma_F$ in the lemma as a product of $I^J$,
$I^{mn}$, $I^{\ell mn}$ and $I^{lmnp}$. We will prove this by induction
on $r$. It is trivial for $r=1$.

Note that the matrix $F^J$ corresponding to $I^J$ is $f_{ij}=J_iJ_j$. The
matrix $F^{mn}$ corresponding to $I^{mn}$ equals 0 everywhere, except
for $f_{mn}=f_{nm}=1$. The matrix $F^{lmn}$ corresponding to $I^{lmn}$
equals zero everywhere except $f_{ln},f_{nl},f_{mn},f_{nm}$. The matrix
$F^{lmnp}$ equals zero everywhere except for $f_{lm},f_{ln},f_{ml},
f_{mp},f_{pm},f_{nl},f_{np}$ and $f_{pn}$.
Finally, $\sigma_{F''}=\sigma_{F'}\sigma_{F}$, where
$$f''_{ij}=f_{ij}+f'_{ij}+\sum_{\ell=1}^r
f_{i\ell}f'_{\ell j}k_\ell\sp({\rm mod}\sp 2).\eqno(3.13)$$

Suppose first that some row or column of $F$ -- say the $i$th -- is
identically zero. Then by (i) or $(3.12a$), respectively, the $i$th
column or row will also be identically zero.
By (3.11) $\sigma$ fixes the $i$th component of each $a$, and so is
the tensor product of $\A^{k_i}$ with some $g^{r-1}_{\hat k}$ permutation
invariant, where ${\hat k}=(k_1,\ldots,k_{i-1},k_{i+1},\ldots,k_r)$. By
the induction hypothesis we are done.

Suppose some $f_{ii}=1$. Let $J_j=f_{ij}$, then $J^2\equiv 2$ (mod 4),
by (ii), so $I^J$ is a permutation invariant. From (3.13) and (ii) we find
that $f(\sigma_J\sigma)_{ij}=0$ for all $j$ -- by the previous paragraph
we are done.

Thus it suffices to assume all $f_{ii}=0$.
Now suppose some $k_i\equiv 2$ (mod 4). Then choose as a simple current
$J_j=\delta_{ij}$. Then by (3.13), $f(\sigma_J\sigma)_{ii}=1$ so by the
preceding paragraph we are done. For $a=0,1,2,3$ let $n_a$ be the number
of $k_i\equiv a$ (mod 4). Then we may assume $n_2=0$. If $n_1=n_3=0$, \ie
if all $k_i\equiv 0$ (mod 4), then products of various $I^{mn}$ with $\sigma$
results in
an $F$-matrix identically zero, by (i). Also, if exactly one $k_i$ is odd, \ie
$n_1+n_3=1$, then
by (ii) the $i$th column of $F$ will be identically zero, and again we are
done.

Suppose $k_i\equiv k_j\equiv \pm 1$ (mod 4), and $f_{ij}=0$, for $i\ne j$.
Then choosing $J_\ell=\delta_{\ell i}+\delta_{\ell j}$ produces $f''(\sigma_J
\sigma)_{ii}=1$, so we are done. Note that if $n_3\leq 1$ and $n_1\ge 2$
(or $n_3\ge 2$ and $n_1\le 1$), then by (i) and (ii) this is inevitable.
Thus we are reduced to two cases: either $n_1,n_3\ge 2$; or $n_1=n_3=1$.

Consider now the first case. Choose any $k_i\equiv k_j\equiv +1$ (mod 4).
We may assume $f_{ij}=1$. Then again by (i) and (ii), there must be a
$k_\ell\equiv -1$ (mod 4) such that $f_{i\ell}=0$. Choosing any other
$k_m\equiv-1$ (mod 4), we then get from (3.13) that $f(\sigma_{ij\ell m}
\sigma)_{ii}=1$.

Finally,  consider $k_1\equiv+1$, $k_2 \equiv-1$
and $k_3\equiv k_4\equiv\cdots\equiv k_r\equiv 0$ (mod 4), where $f_{ij}=0$
for all $i,j=1,2$.
By (ii) we get $f_{i1}=f_{i2}$ $\forall i$, so by (i), $f_{ij}=f_{ji}$ $\forall
i,j$. If now $f_{ij}=1$
for any $i,j>2$, then hit $I^\sigma$ with $I^{ij}$. Thus we have succeeded
in reducing the proof to showing that the permutation invariants associated
with the matrices
$$F=\left(\matrix{0&0&1&\cdots&1\cr 0&0&1&\cdots&1 \cr 1&1&0&\cdots&0\cr
\vdots&\vdots&\vdots&&\vdots\cr 1&1&0&\cdots&0\cr}\right) \eqno(3.14)$$
can be written as in (3.5). But this
is obvious from (3.13): the product $I^{123}\cdots I^{12r}$ works.

This concludes the proof of Thm.1.

\bigskip\bigskip\noindent{{\bf 4. Simple current invariants}}\bigskip

Throughout this section fix $r$ and $k$ and write $S_{ab}$ for $S^{(k)}_{ab}$,
and $\rho$ for $\rho_r$. Recall the definition of simple current invariant
given in (2.6).

One of the
main reason we are interested in simple current invariants is that
they seem to exhaust most physical invariants. For example, all but finitely
many physical invariants for $A_1$ and $A_2$ are simple current invariants
(or their conjugates).
This seems to remain true to some extent for $g^r$, though for $r>1$ we
have infinitely many ``exceptionals''.
It turns out that simple current invariants have the very pleasant property
that they are nice enough to classify completely \GRS, largely because of the
important relation
$$S_{Ja,J'b}=-1^{J\cdot (b-\rho)+J\cdot(a-\rho)+J\cdot J'}S_{ab}. \eqno(4.1)$$

In this section we classify all the simple current invariants
of $g^r_k$. In particular we will prove:

\medskip \noindent{{\bf Theorem 2.}}\quad {\it Let $M$ be a simple current
invariant of $g^r_k$, where at most three $k_i=2$.
Then there exist simple currents $J^1,\ldots,J^m$ and a permutation
invariant $I^\sigma$ whose $\pi=id.$, such that}

\item{(a)} $J^{i2}\equiv 0\sp$ (mod 4) $\forall i$, {\it and}
$J^i\cdot J^j\equiv 0\sp$ (mod 2) $\forall i,j$;

\item{(b)} $M=\bigl( \prod^m_{i=1} M(J^i)\bigr) I^\sigma$, {\it where} $M(J^i)$
{\it is defined in} $(2.5a$).\medskip

Another characterization of the simple current invariants is given by equation
(4.11) below. The arguments we will use have much in common with the much more
general
Ref.~\GRS, but because we are focusing here on the special case of $g^r_k$,
our conclusion can be made more explicitly, and we are required to assume
a little less (in particular, Ref.~\GRS{} finds all {\it strongly physical}
simple current invariants, while we find here all {\it physical} simple
current invariants). The strange-looking condition that at most three $k_i=2$
is unfortunate,
and is related to the regularity condition of Ref.~\GRS. However, it is
more than sufficient for the applications we have at present (namely,
Sect.7 below, and \Appl). Nevertheless, it is something we would like to
return to in the future.

We will begin our proof of Thm.2 by proving the following general lemma.
This lemma will be used throughout this paper, and
is clearly of importance in its own right. It holds in a far greater
context than just $g^r_k$. To prove it, we will only need to assume
(P1)-(P3).

\medskip\noindent{\bf Lemma 2.}\quad {\it Let $M$ be any physical invariant
(not
necessarily a simple current invariant). Let $\I_L,\I_R\subset\I$ be defined
by $J\in\I_L$ iff $M_{J\rho,\rho}\ne 0$, $J'\in\I_R$ iff $M_{\rho,J'\rho}
\ne 0$. Define $\p_L=\{a\in P^r_k\,|\, \exists b$ for which $M_{ab}\ne 0
\}$, $\p_R=\{b\in P^r_k\,|\, \exists a$ for which $M_{ab}\neq 0\}$. Then}

\item{(i)} {\it $\I_L$ and $\I_R$ are subgroups (hence subspaces over $\Z_2$)
 of $\I$;}

\item{(ii)} $M_{ab}\ne 0$ {\it implies} $J\cdot(a-\rho)\equiv J'\cdot(b-\rho)
\equiv 0$ (mod 2) $\forall J\in\I_L,J'\in\I_R$;

\item{(iii)} $M_{ab}=M_{Ja,J'b}$ $\forall J\in\I_L,J'\in\I_R$;

\item{(iv)} $J^2\equiv 0\sp$ (mod 4), $J\cdot J'
\equiv 0\sp$ (mod 2) $\forall J,J'\in\I_L$ ({\it similarly for} $\I_R$).

\item{(v)} {\it Suppose in addition that $M_{a\rho}\neq 0$ iff $a\in\I_L\rho$.
 Then}
$$\eqalignno{a\in \p_L\sp\sp{\rm iff}&
\sp\sp J\cdot(a-\rho)\equiv 0\qquad ({\rm mod}\sp 2),
\qquad\forall J\in\I_L;&(4.2a)\cr
\|\I_L\|=&\sum_b M_{\rho b}S_{\rho b}/S_{\rho\rho}.&(4.2b)\cr}$$

To prove this, we begin with the calculation (from $M=S^{\dag}MS$ and (4.1))
$$\eqalignno{M_{J\rho,\rho}=&\sum_{a,b}S_{J\rho,a}M_{ab}S_{b\rho}=\sum_{a,b}
S_{\rho a}M_{ab}S_{b\rho}\cdot -1^{J\cdot(a-\rho)}&\cr
\leq& \sum_{a,b} S_{\rho a}M_{ab}S_{b\rho}=M_{\rho\rho}=1,&(4.3)\cr}$$
for any $J\in\I$. Thus $J\in\I_L$ iff $J\cdot(a-\rho)\equiv 0$ (mod 2), for
all $a\in \p_L$. This proves (i),(ii). (iii) follows now from a
calculation similar to (4.3), using (ii). (iv) comes from $T$-invariance,
applied to $M_{J\rho,\rho},M_{JJ'\rho,\rho}\ne 0$.

To prove (v), look at the equation $MS=SM$: for any $a$ such that $(a-\rho)
\cdot \I_L\equiv 0$ (mod 2), it implies
$$\sum_{b}M_{ab}S_{b\rho}=\sum_b S_{ab}M_{b\rho}=\sum_{J\in\I_L} S_{a\rho}
\cdot -1^{J\cdot(a-\rho)}=\|\I_L\|S_{a\rho},\eqno(4.4)$$
the second equality using (iii).
But each $S_{b\rho}>0$. This gives us $(4.2a)$. $(4.2b)$ follows immediately
by restricting (4.4) to $a=\rho$. This completes the proof of the lemma.

Let us begin by deriving some useful expressions.
We are interested for the remainder of this section
 in $M$ being a simple current invariant. Then $(4.2b)$
and (iii) tell us that the cardinalities $\|\I_L\|=\|\I_R\|$ are equal.
For any $a,c\in\p_k^r$, using $SM=MS$ and equations
(2.6) and (4.1) we get the important
$$\sum_{Jc\in \I c} M_{Jc,c}\cdot -1^{J\cdot(a-\rho)}=\sum_{Ja\in\I a}
M_{a,Ja}\cdot -1^{J\cdot(c-\rho)},\eqno(4.5)$$
which holds whenever $S_{ac}\ne 0$. Choosing $a\in\p_L$ and $c=\rho$ tells us
$$\|\I_L\|=\sum_{Ja\in\I a} M_{a,Ja}=a{\rm th
\sp row \sp sum \sp of\sp}M,\eqno(4.6)$$
with a similar expression for the column sum.

Now, choose any $a\in {\cal P}_L$, $J_R\in{\cal J}_R$. Then $M_{a,J_aa}=
M_{a,J_RJ_aa}\ne 0$ for some $J_a\in {\cal J}$. $T$-invariance tells us both
$$\eqalignno{J_a^2\equiv \,&2 J_a\cdot(a-\rho)\sp\sp
({\rm mod}\sp 4),&(4.7a)\cr
J_R\cdot(a-\rho)\equiv\,& J_a\cdot J_R\sp\sp({\rm mod}\sp 2).&(4.7b)\cr}$$

For any $a$, let $F_a$ ($F^a$) denote the number of $J\in {\I}_L$
($J\in{\I}_R)$ such that $Ja=a$. $a$ is called a {\it fixed point} of
${\I}_L$ if $F_a>1$. A fixed point has the (necessary but not
sufficient) property that some of its components $a_i$ must equal $k_i'/2$.

Say $a\in \p_L$ has {\it property} ($*_L$) if there exists a $J_a\in \I$
with the property that $M_{ab}\ne 0$ iff $b\in \I_RJ_a a$. Similarly, say
$c\in \p_R$ has property $(*_R)$ if $\exists J^c\in\I$ such that $M_{bc}\ne 0$
iff $b\in \I_LJ^cc$. We would like to show that every $a\in \p_L$ has
property $(*_L)$, and every $c\in\p_R$ has propery $(*_R)$. Note that,
because of Lemma 2(iii) and (4.6), if $a$ has property $(*_L)$ then $M_{ab}$
will either equal $F^a$ or 0.

Consider first $a\in{\cal P}_L$ which is not a fixed point of $\I_R$. Then
from (4.6) and Lemma 2(iii), $a$ must have property $(*_L)$. Similarly, any
$c\in\p_R$  which is not a fixed point of $\I_L$ must have propery $(*_R)$.

On the other hand, the treatment of the fixed points of $\I_R$, or of
$\I_L$, is a little more complicated, and in fact because of these
complications we will have to assume that only a few $k_i=2$.

Suppose $c$ is not a fixed point of $\I_L$, but $a$ is a fixed point of
$\I_R$, and suppose $S_{ac}\ne 0$. Then (4.5) reduces to
$$\|\I_L\|\cdot -1^{J^c\cdot(a-\rho)}=\sum_{Ja\in\I a}M_{a,Ja}\cdot
-1^{J\cdot(c-\rho)}.\eqno(4.8a)$$
Because the $a$th row sum of $M$ equals $\|\I_L\|$, the only way equality
can hold in (4.$8a$) is if the signs are all equal, \ie if
$$J^c\cdot(a-\rho)\equiv J\cdot(c-\rho)\sp({\rm mod}\sp 2)\eqno(4.8b)$$
whenever $M_{a,Ja}\ne 0$.

Choose any $J',J''\in\I$ such that $M_{a,J'a}\ne 0$ and $M_{a,J''a}\ne 0$,
and let $J=J'J''$. Then $(4.8b)$ becomes
$$J\cdot(c-\rho)\equiv 0\sp({\rm mod}\sp 2). \eqno(4.8c)$$
This holds for any $c\in {\cal P}_R$ such that $c$ is not a fixed point
of $\I_L$, and $S_{ac}\ne 0$. We are free to demand $J'_i=J''_i=0$, and
hence $J_i=0$, for any $i$ for which $a_i=k_i'/2$.

Define $\I^a=\{J\in \I\,|\,J_i=0$ when $a_i=k_i'/2\}$, so $J',J'',J\in\I^a$.
Let $\I^a_R$  be the projection of $\I_R$ into that space, \ie
$\I_R^a=\{J\in\I^a\,|\,\exists
J_R\in\I_R$ such that $J_{Ri}=J_i$ when $a_i\ne k'_i/2\}$. Then $\I^a$ is a
subspace of $\I$, and $\I^a_R$ is a subspace of $\I^a$. Let $(\I^a_R)^\perp
\subset \I^a$ denote the space
$$(\I^a_R)^\perp=\{x\in \I^a\,|\,\sum_{i=1}^rx_i{\tilde J}_i\equiv 0
\sp\sp({\rm mod}\sp 2),\sp\sp\forall \tilde J\in\I^a_R\}.\eqno(4.9a)$$
To any $x\in(\I^a_R)^\perp$ assign the weight $c=c(x)$ by $c_i=1$ if
$x_i\equiv 0$, and
$c_i=2$ if $x_i\equiv 1$. Then $x\in\I^a$ implies $S_{ac}\ne 0$, and
$x\in(\I^a_R)^\perp$ implies $c\in {\cal P}_R$. Moreover, provided no $k_i=2$,
$c$ will not be a fixed point of $\I_L$. We will return shortly to the
case where some $k_i=2$, but for now assume none do.

Putting this $c$ in $(4.8c$) is equivalent to noting that
$$J\in\bigl((\I_R^a)^\perp\bigr)^\perp=\{\tilde J\in\I^a\,|\,\sum_{i=1}^r
{\tilde J}_ix_i\equiv 0\sp\sp({\rm mod}\sp 2),\sp\sp\forall x\in
(\I^a_R)^\perp\}=\I^a_R,\eqno(4.9b)$$
\ie $\exists J^0\in\I_R$
such that $J_i=J^0_i$ when $a_i\ne k'_i/2$. Thus $J''a=J^0J'a$.

This shows that even fixed points $a\in{\cal P}_L$ obey $(*_L)$
(just choose $J_a=J'$). A similar argument holds for $(*_R)$. But this
argument assumed no $k_i=2$.

And what if some $k_i=2$? Let $t$ be the number of $k_i=2$. Let $t(a)$
be the number of $i$ such that  $a_i=2=k_i$. Recall the map $x\mapsto c(x)$,
taking $(\I^a_R)^{\perp}$ into $\p_R$. Suppose, for a given $a\in
{\cal P}_L$, we know that every $x\in(\I_R^a)^\perp$ is such that $c(x)$
obeys property $(*_R$). Then by the above argument we get that $a$ obeys
$(*_L)$. Now, $c=c(x)$ will necessarily obey $(*_R$) if $t(c)=0$ or 1,
because then $c$ could not be a  fixed point. Therefore $a\in\p_L$ will
necessarily obey $(*_L)$, if either $t(a)=t$ or $t-1$. Together these two
statements tell us that $(*_R)$
 is satisfied whenever $c=c(x)$ has $t(c)=0,1,t-1$ or $t$.

So for any $t\le 3$, \ie when there are at most 3 $k_i=2$,
we get that all $a\in {\cal P}_L$ satisfies
$(*_L)$. The same argument applies to $\p_R$.

Equation (4.5) now collapses to the equation
$$J^c\cdot(a-\rho)\equiv J_a\cdot(c-\rho)\sp\sp({\rm mod}\sp 2) \eqno(4.10a)$$
whenever $a\in {\cal P}_L$, $c\in {\cal P}_R$, and $S_{ac}\ne 0$. From
a similar to argument to that used in equations $(4.8c$)-(4.9$b$) above, we
get that if $a\equiv a'$ (mod 2), and $a\in
{\cal P}_L$, then we may choose $J_a=J_{a'}$. (Put $J=J_aJ_{a'}$; we may
demand $J_i=0$ whenever either $a_i=k'_i/2$ or $a_i'=k'_i/2$, etc.)
Hence, provided no $k_i=2$ (we will return shortly to the case where some
$k_i=2$),
we may drop the irritating condition $S_{ac}\ne 0$ from (4.10$a$). (4.10$a$)
also
tells us that we may choose $J_a=J_{a'}J_{a''}$, when $a-\rho\equiv(a'-\rho)
+(a''-\rho)$ (mod 2) and $a',a''\in{\cal P}_L$.

However by definition, $J^c\in \I_RJ_b$ for $b=J^cc$. Hence, using $(4.7b$),
we may write $(4.10a$) in the more convenient form
$$J_b\cdot(a-\rho)+J_a\cdot(b-\rho)\equiv J_a\cdot J_b\sp\sp
({\rm mod}\sp 2)\sp \forall a,b\in{\cal P}_L.\eqno(4.10b)$$

Once again we must return to the case where there are $k_i=2$. Here however
we do not require any restriction on the size of $t$ (=$\,$the number of
$k_i=2$),
provided we know $(4.10a)$ is satisfied whenever $S_{ac}\ne 0$, and $a\in
\p_L$, $c\in\p_R$. The idea is
that we are completely free to choose the values of $(J_a)_i$ whenever
$a_i=2=k_i$, so we will try to choose them so that (4.10) holds. The argument
is not difficult and we will give only a sketch. Note first that if $t(a)=0$,
then the above argument remains valid and the condition ``$S_{ac}\ne 0$''
may still be dropped; similarly we can take $J_a=J_{a'}$ if $a\equiv a'$ (mod
2) and $t(a)=0$.

Define the projection $p(a)\in \I$ by $p(a)_i\equiv a_i$ (mod 2) if $k_i=2$,
and $p(a)_i\equiv 0$ if $k_i\ne 2$. Begin by finding a set of $a^\alpha
\in{\cal P}_L$ for which $p(a^\alpha-\rho)$ forms a basis of $p({\cal P}_L
-\rho)$; for convenience make it ``upper triangular''. Then, by choosing  the
free values of $(J_{a^\alpha})_\ell$ appropriately, it is possible to force
$(4.10b)$ to be satisfied for any $a=a^\alpha$, $b=a^\beta$. Now, any
$a\in{\cal P}_L$ can be written
as $a-\rho\equiv\sum m_\alpha(a^\alpha-\rho)+a'-\rho$ (mod 2), for some
$m_\alpha\in\Z$ and $a'\in{\cal
P}_L$ with $t(a')=0$ and $a'_i\ne k'_i/2$ for all $i$. Define
$J_a'=\sum m_\alpha J_{a^\alpha}+J_{a'}$. Then the usual
argument shows that $J=J'_aJ_a$ satisfies $(4.9b$), so we are free to
choose $J_a=J_a'$. In this way we get that $(4.10b$) is satisfied for
all $a,b\in{\cal P}_L$.

Let us summarize our results so far. What specifies our simple current
invariant $M$? Choose two subgroups $\I_L,\I_R\subset\I$, with equal orders
$\|\I_L\|=\|\I_R\|$. For each $a\in{\cal P}_L$ find a $J_a\in\I$. Define
$M$ by
$$M_{ac}=\left\{\matrix{F^a&{\rm if\sp}a\in {\cal P}_L\sp{\rm and}\sp c\in
\I_RJ_aa\cr 0&{\rm otherwise}\cr}\right. .\eqno(4.11)$$
Both $\I_L$ and $\I_R$ must satisfy Lemma 2(iv). Also,
in order for $M$ to be a modular invariant, equations ($4.10b$) and $(4.7$)
must be satisfied. There are various consistency conditions which must also
be satisfied (\eg $\I_La=\I_Lb$ iff $\I_R J_a a=\I_RJ_bb$), but these will
be automatically satisfied if these other conditions are.

Our remaining
task is to {\it solve} these conditions, \ie to show the matrix $M$ in
(4.11) satisfying these conditions must be as in Thm.2. To this
effect, note that if we had $\I_L=\I_R$, and in addition each $J_a=0$,
then we may write $M$ as
$$M=\prod_{J\in\beta} M(J),\eqno(4.12)$$
where $\beta$ is any basis of $\I_L$ and $M(J)$ denotes the simple current
invariant defined in equation
(2.5$a$). This can be proven by explicitly expanding
(4.12), using the fact that $J_L\cdot J_L'\equiv 0$
$\forall J_L,J_L'\in\I_L$.

So our task is to find a permutation invariant $I^\sigma$ (3.11), whose
conjugation $\pi=id.$, such that $I^\sigma$ takes $\I_R\rho$ to
$\I_L\rho$ and each $J_aa$ into $\I_La$. Inspecting (3.9), we may expect
to take as $F$ the transpose of the matrix which sends $a-\rho$ to $J_a$.
This is indeed the right idea, though this matrix is only defined for
$a\in{\cal P}_L$.

Let $n=\|\I_L\|=\|\I_R\|$. Find $a^i\in\p_L$, $i=1,\ldots,r-n$, so that the
$a^i-\rho$ form a basis
for the vector space spanned by ${\cal P}_L-\rho$ (mod 2).
It is possible to choose bases
$\{J_L^j\}$ of $\I_L$ and $\{J_R^j\}$ of $\I_R$, and a set of vectors
$\{b^j\}$, for $j=1,\ldots,n$, such that $b^i\cdot J^j_L\equiv b^i\cdot
J^j_R\equiv \delta_{ij}$ (mod 2). (As before, dot products between $b$'s, $c$'s
and $J$'s will look like $\sum b_iJ_i$, and those between two $J$'s will be
like $\sum J_iJ'_ik_i$.) Then the $\{a^i-\rho\}$ and $\{b^j\}$
together form a basis for all of $\I$. Define $M_{ij}=J_{a^i}\cdot b^j$.

Finally, define $J_c'$ linearly for all $c\in\I$, by putting
$$\eqalignno{J'_{a^i-\rho}=&\,J_{a^i}+\sum_{j=1}^nM_{ij}J^j_R,\sp i=1,\ldots,
r-n,&(4.13a)\cr
J'_{b^j}=&\,0,\sp j=1,\ldots,n.&(4.13b)\cr}$$
Hence for all $a\in \p_L-\rho$ and all $b\in$ span of $b^j$, we have
$J'_a\cdot b\equiv 0$ (mod 2) and $J_b'=0$. Also, $J_a$ is only defined
mod $\I_R$, so
we may replace $J_a$ with $J'_{a-\rho}$ without affecting the $M$ in
(4.11).

An easy calculation now shows
$$\eqalignno{J_d'\cdot c+J_c'\cdot d\equiv &\,J_c'\cdot J_d'\sp
\sp({\rm mod}\sp 2),&(4.14a)\cr
J_c'{}^2\equiv &\,2J_c'\cdot c\sp\sp({\rm mod}\sp 4),&(4.14b)\cr}$$
$\forall c,d\in\I$. Define a matrix $F=(f_{ij})$ of 0's and 1's by $(J_c)_i
=\sum_{j=1}^r f_{ij}c_j$. Then (4.14) becomes equations (3.12), so $\sigma_F$
is a permutation
invariant. Also, $\sigma_F(J_{a-\rho}'a)=a$ for all $a\in {\cal P}_L$:
$$\eqalign{(J^{J'_{a-\rho}a})_i=&\sum_j f_{ji}(J'_{a-\rho}a-\rho)_j=\sum_j
f_{ji}((J'_{a-\rho})_jk_j+a_j-1)\cr
=&\sum_jf_{ji}(\sum_\ell f_{j\ell}(a_\ell-1)k_j+a_j-1)=
\sum_\ell (a_\ell-1) f_{i\ell}=(J'_{a-\rho})_i,\cr}$$
so $\sigma_F(J'_{a-\rho}a)=J'_{a-\rho}(J'_{a-\rho}a)=a$. This means
the set ${\cal P}_R^F$ for the simple current invariant $M^F=MM^{\sigma_F}$
equals the set ${\cal P}_L={\cal P}_L^F$, which implies $\I_R^F=\I_L=\I_L^F$.
Also, the $J^F_a$ for this invariant can all be taken to be 0.
This then concludes the proof of Thm.2.

We know of no examples of simple current invariants of $g^r_k$, where there
are more than 3 $k_i=2$, which does not satisfy (i) and (ii) of Thm.2.
To remove the condition on the number $t$ of $k_i=2$ in Thm.2, it suffices to
prove that for all $c\in\p_R$ with $c_i=1,2$, there exists a $J^c\in\I$
satisfying: $M_{bc}\ne 0$ iff $b\in\I_LJ^cc$. In this section we have only
proven it for $c$ with $t(c)=0,1,t-1$ or $t$, where $t(c)$ is the number of
$i$ with $c_i=k_i=2$. That is the source of our restriction $t\le 3$.

\bigskip\bigskip\bigskip\bigskip
\noindent{{\bf 5. Exceptional invariants}}\bigskip

(P1)-(P3) are not the only properties a physically acceptable partition
function must satisfy.
We learn in \MS{} that it will look like
$$\eqalignno{Z=&\sum_{i=1}^\alpha ch_i ch_{\tau i}'{}^*&(5.1a)\cr
{\rm where}\sp & ch_i=\sum_{a\in P^r_k} m_{ia}\chi_{a}^{k},\sp
\sp ch_j'=\sum_{b\in P^r_k} m_{jb}'\chi_{b}^{k},&(5.1b)\cr}$$
where $m_{ia},m_{jb}'$ are non-negative integers and $\tau$ is some
bijection. The $ch_i,ch_j'$ are characters for the LHS, RHS maximally
extended chiral algebras ${\cal C}_L$, ${\cal C}_R$. We may label them
so that $m_{i\rho}=\delta_{i1}$, $m_{j\rho}'=\delta_{j1}$. The characters
$ch_i,ch_j'$ transform modularly with unitary symmetric matrices $S^{e}_{ij}
,T^{e}_{ij},S^{e}_{ij}{}',T^{e}_{ij}{}'$, and satisfy
$$S^{e}_{1i}\ge S^{e}_{11}>0,\sp S^{e}_{1j}{}'\ge S^{e}_{11}{}'>0,
\sp\sp\forall i,j.\eqno(5.1c)$$

In this section,
by a {\it strongly physical invariant} $M$ we mean a physical invariant
satisfying equations (5.1). There are other conditions $Z$ and its chiral
algebras must obey in order to be physically acceptable -- \eg the fusion rules
of its chiral algebras must be non-negative integers.
In Sect.7 another condition will be introduced.

Each simple current invariant is
strongly physical, and defines chiral extensions ${\cal C}_L={\cal C}(\I_L)$,
${\cal C}_R={\cal C}(\I_R)$ of the affine algebra. The simple current
invariants and their conjugations can be thought of as the regular series
of physical invariants for $g^r_k$, generalizing the ${\cal A}_k$, ${\cal
D}_k$ series of $A_1$. Any other physical invariants can be called
{\it exceptional}. An exceptional invariant comes in two basic kinds: either
it corresponds to an
exceptional automorphism $\tau$ of simple current extensions (\eg
${\cal E}_{16}$ for $A_1$); or one of
its (maximally extended) chiral algebras ${\cal C}_L$, ${\cal C}_R$ are
exceptional, \ie at least one is not a simple current extension (\eg
${\cal E}_{10}$ and ${\cal E}_{28}$ of $A_1$).
We will call exceptionals of the first kind ${\cal E}_{16}$-like,
and those of the second kind ${\cal E}_{10}$-like.

In this section we
investigate the nature and existence of both these kinds of
exceptional invariants for $g^r_k$, aided by the powerful machinery of
Ref.{} \MS. We can expect in our WZNW classifications that ``almost every''
level $k$ will be nicely behaved; our goal in this section is to find these
generic results and to isolate those levels where sporatics may occur.
We will be able to say much more here about the ${\cal E}_{16}$-like
exceptionals, than about the ${\cal E}_{10}$-like exceptionals. The latter
are most easily handled using additional machinery \STA.

The main examples of chiral extensions are due to simple currents. By a
simple current chiral extension ${\cal C}(\I_L)$ we simply mean one whose
characters, when projected down to the affine variables, look like
$$ch_{[a]_i}={G([a]_i)\over F_a}\sum_{J\in \I_L}\chi_{Ja},\quad i=1,
\ldots,f(a),\eqno(5.2a)$$
where as usual $F_a$ is the number of $J\in\I_L$ with $Ja=a$, where
$a\in\p_L$, and where $[a]$ denotes the orbit $\I_La$. The coefficients
$G([a]_i)$ are positive integers satisfying
$$\sum_{i=1}^{f(a)} \bigl( G([a]_i)\bigr)^2=F_a.\eqno(5.2b)$$
When $f(a)=1$ we will often write $ch_{[a]_1}$ as $ch_{[a]}$.
We are most familiar with the case where all $G([a]_i)=1$, in which case
$f(a)=F_a$, but we will see that it is important to allow $G>1$ as well.
We will have more to say about this shortly. Equations $(5.2$) tell us that
the invariant $\sum |ch_{[a]_i}|^2$ will equal the simple current invariant
in (4.12); indeed they are the only sets of characters obeying (5.1) with
that property. We will usually use primes to denote the corresponding
quantities for ${\cal C}_R={\cal C}(\I_R)$, \eg $[b]'=\I_Rb$.

Equations (5.2) give us
$$\sum_{i=1}^{f(a)}G([a]_i)\,S^e_{[a]_i,[b]j}={\|\I_L\|\,G([b]_j)\over F_b}
S_{ab},\eqno(5.3a)$$
where $S_{ab}=S^{(k)}_{ab}$ is the $S$-matrix for $g^r_k$.
Equation $(5.3a)$ fixes $S^e_{[a]_i,[b]_j}$ if either $f(a)=1$ or $f(b)=1$.
Formally defining the fusion rules $N^e_{[a]_h,[b]_i,[c]_j}$ by Verlinde's
formula, we get from $(5.3a)$ that if $f(a)=f(b)=1$, then
$$N^e_{[a],[b],[c]_j}={G([c]_j)\over G([a])\,G([b])}\sum_{d\in\I_Lc}N_{abd}.
\eqno(5.3b)$$
The remaining values of $N^e$ and $S^e$ for simple current extensions are
harder to compute in general, though $(5.3a$) does tell us
$$\sum_{h=1}^{f(a)}\sum_{i=1}^{f(b)}G([a]_h)\,G([b]_i)\,
N^e_{[a]_h,[b]_i,[c]_j}=G([c]_j)\sum_{d\in\I_Lc} N_{a,b,d}.\eqno(5.3c)$$
We can see from these equations that some $G([a]_i)\ne 1$ will often lead
to non-integer fusion rules. Indeed, it is easy to show that for $r=2$,
integer fusion rules require that all $G([a]_i)=1$. However, this is
not true for all $r$ -- see (5.12$b$) for a counter-example when $r=8$.

Apparently, the constraints of unitarity,
symmetry, the relations $S^{e2}=(S^eT^e)^3=C^e$, and integrality of the
fusion rules {\it do not} suffice to essentially
fix $S^e$, and hence ${\cal C}(\I_L)$, for a given $\I_L$. Again, see
(5.12$b$) for a counter-example. This seems like an important observation,
since it is not hard to show that for any simple current extension
${\cal C}(\I_L)$, any physical invariant $M^e$ of ${\cal C}(\I_L)$ will
produce a physical invariant $M$ of $g^r_k$, through the formula
$$M_{ab}=\sum_{i=1}^{f(a)}\sum_{j=1}^{f(b)}G([a]_i)\,G([b]_j)\,
M^e_{[a]_i,[b]_j}.\eqno(5.3d)$$
All that is needed to prove that $(5.3d)$ is a physical invariant is the
unitarity and symmetry of $S^e$, along with $(5.3a$).
For a given $\I_L$ there will be several such $S^e$. Indeed, the $g^2_{4,4}$
exceptional is a realization of this. It corresponds to a symmetry of a
simple current chiral extension with all $G([a]_i)=1$ except $G([33])=2$; the
matrix $S^e$ is real and obeys $(S^eT^e)^3=I$, as it should. However,
$N^e_{[33],[33],[33]}=1/2$. The $g^2_{4,4}$ exceptional can also be interpreted
as the diagonal invariant for an exceptional chiral extension, but again
the fusion rules will not be integers \FKSV.

We will begin by giving a characterization of ${\cal E}_{10}$-like
exceptionals.

\medskip\noindent{{\bf Theorem 3.}}\quad {\it Let $M$ be any strongly physical
invariant of $g^r_k$, where at most three $k_i=2$ and at most one $k_i=4$.
Suppose $M_{a\rho}\ne 0$ iff $a\in\I_L\rho$.
Then its LHS chiral algebra ${\cal C}_L$ will be a simple current chiral
extension ${\cal C}(\I_L)$.}\medskip

\noindent{\it Proof.}\quad Define $Z'$ by
$$Z'=\sum_{i=1}^\alpha|ch_{i}|^2=\sum_{i=1}^\alpha\sum_{a,b\in P_k^r}m_{ia}
m_{ib}\chi_{a}\chi_{b}^*,\quad i.e.\sp M'_{ab}=\sum_{i=1}^\alpha
m_{ia}m_{ib}.\eqno(5.4a)$$
Then $M$ strongly physical implies $M'$ will be a physical invariant.
$M'$ will have left and right chiral algebras equal to ${\cal C}_L$. It is
more
convenient to work with $M'$ than with $M$. Our main tools will be Lemma 2
and Perron-Frobenius theory \MAT.

Write $M'$ as a direct sum of indecomposable submatrices $B_\ell$, $\ell=
1,\ldots,\beta$. Let $W_\ell\subset \p_L$ be the weights contained in the block
$B_\ell$.
Call $W_1$ the unique block containing $\rho$. The Perron-Frobenius
eigenvalue of $B_1$ is $r(B_1)=\|\I_L\|$, so by Lemma 3 of \GR{} $r(B_\ell)
\le \|\I_L\|$ for each $\ell$. Consider now any block $B_\ell$ with some
$a\in W_\ell$ having $F_a=1$. Defining
$$(B_\ell^a)_{bc}=\left\{\matrix{1&{\rm if}\sp b,c\in\I_La\cr
0&{\rm otherwise}\cr}\right. ,\eqno(5.4b)$$
we see that $B_\ell\ge B^a_\ell$, by Lemma 2 and $M'_{aa}\ge 1$,
and hence that $r(B_\ell)\ge r(B^a_\ell)$, with equality iff $B_\ell=
B_\ell^a$ \MAT. But $r(B^a_\ell)=\|\I_L\|$. Therefore whenever $F_a=1$,
there exists a unique $\ell(a)$ such that $m_{i,a}=\delta_{i,\ell(a)}$,
and $W_{\ell(a)}=\I_La$.

Note from Lemma 2 that $M'_{aa}=M'_{a,Ja}=M'_{Ja,Ja}$ for any
$J\in\I_L$. Rewriting this in terms of the $m_{ib}$, the triangle inequality
tells us $m_{ia}=m_{i,Ja}$, for all $1\le i\le \alpha$, $a\in\p_L$, and
$J\in\I_L$.

{}From $MS=SM$, we get that
$$\sum_{b\in W_\ell}(B_\ell)_{ab}S_{b\rho}=\sum_{b\in W_1}
S_{ab}M'_{b\rho}=\|\I_L\|S_{a\rho},\eqno(5.5a)$$
for all $a\in W_\ell$. In other words, the vector $x_\ell$ defined
by $(x_\ell)_a=S_{a\rho}$, for $a\in W_\ell$, is the Perron-Frobenius
eigenvector of $B_\ell$, and $r(B_\ell)=\|\I_L\|$. Now consider $B^\infty
={\rm lim}_{n\rightarrow\infty} (M'/\|\I_L\|)^n$. This limit will exist, and
in fact will be the direct sum of the matrices $B^\infty_\ell$ defined by
$$(B^\infty_\ell)_{ab}={(x_\ell)_a (x_\ell)_b\over \sum_{c\in W_\ell}
(x_\ell)_c^2}=C_\ell S_{a\rho}S_{b\rho},\quad {\rm where}\sp
C_\ell={1\over \sum_{c\in W_\ell} S_{c\rho}^2},\eqno(5.5b)$$
and $a,b\in W_\ell$. $M^\infty$ will be a modular invariant. This means
$$C_i\sum_{b\in W_i} S_{a\rho}S_{b\rho}S_{bc}=C_j\sum_{b\in W_j}
S_{ab}S_{b\rho}S_{c\rho},\quad \forall a\in W_i,c\in W_j. \eqno(5.5c)$$
Consider the special case where $F_a=1$. Then ($5.5c$) simplifies to the
statement: the ratio $S_{ab}/S_{b\rho}$ will be independent of whichever
$b\in W_j$ is chosen, for any fixed $j$.

For each $\ell=1,\ldots,r$ with $k_\ell\neq 1,4$, define $(f^\ell)_j=1+2
\delta_j^\ell$. Then $f^\ell\in \p_L$ and $F_{f^\ell}=1$. For any $b,c
\in W_j$, for any $j$, we then have
$${S_{f^\ell b}\over S_{b\rho}}={S_{f^\ell c}\over S_{c\rho}},\quad
i.e.\sp {\sin(3\pi b_\ell/k'_\ell)\over\sin(\pi b_\ell/k'_\ell)}=
{\sin(3\pi c_\ell/k'_\ell)\over\sin(\pi c_\ell/k'_\ell)}.\eqno(5.5d)$$
But $\sin(3x)/\sin(x)=3-4\sin^2(x)$, so $\sin(\pi b_\ell/k_\ell')=\pm
\sin(\pi c_\ell/k'_\ell)$, \ie $c_\ell=b_\ell$ or $k_\ell'-b_\ell$.
This conclusion is automatic when $k_\ell=1$. Finally, $T$-invariance forces
it for $k_\ell=4$ (provided there is only one such $k_\ell$).

Thus, $M'$ is a simple current invariant. From the analysis of Sect.4, we
get that $M'$ can be written in the form of (4.11), for $\I_L=\I_R$ and
all $J_a=0$. Therefore $\sum_i m_{ia}^2=F_a$. Also, for each $i=1,\ldots,
\beta$ there is a $a^i\in\p_L$ such that $W_i=\I_La^i$.  \qquad QED\medskip

What this tells us is that in order for $M$ to be ${\cal E}_{10}$-like,
there must be some weight $a\not\in \I\rho$, such that either $M_{\rho a}
\neq 0$ or $M_{a\rho}\ne 0$. That something strange can happen when two
$k_i=4$, is shown by the $g^2_{4,4}$ exceptional.

Next, we will investigate the existence of ${\cal E}_{16}$-like exceptionals.
The following theorem says that they only exist for certain small $k_i$.
This is one of the main results of this paper.

\medskip\noindent{{\bf Theorem 4.}}\quad {\it Let $M$ be a strongly physical
invariant of $g^r_k$, where no $k_i=2$ and at most one $k_i=4$. Assume
$M_{a\rho}\ne 0$ iff $a\in \I_L\rho$, and $M_{\rho b}\ne 0$ iff $b\in \I_R
\rho$. Suppose for any $k_i=4$, $8$, $12$ or $16$ that there does not exist a
$J'\in\I_R$ with $J'_j=\delta_{ij}$, and for any $k_i=4$ or $8$ that there
does not exist a $J\in\I_L$ with $J_j=\delta_{ij}$. Then there exists a
permutation invariant $I^\sigma$ such that }
$$M=M(J^1)\cdots M(J^m)\cdot I^\sigma,\eqno(5.6)$$
{\it for some basis $\{J^1,\ldots,J^m\}$ of $\I_L$}.\medskip

\noindent{{\it Proof.}}\quad Thm.3 tells us $M$ can be thought of as a
bijection $\tau$ between simple current extensions ${\cal C}(\I_L)$ and
${\cal C}(\I_R)$. Our proof will follow as closely as possible the proof
of Thm.1.

For each $i=1,\ldots,r$ with $k_i\ne 1$, define $(f^i)_j=1+2\delta_j^i$. Then
$f^i\in\p_L$, and $F_{f^i}=1$. Our first task will be to show $[b^i]'_j
=\tau([f^i])$ is not a fixed point of $\I_R$.

Suppose for contradiction that $b^i$ is a fixed point of $\I_R$, and call
$F'_i=F'_{b^i}$. Then
$$S_{\rho f^i}={1\over F'_i}S_{\rho b^i}.\eqno(5.7a)$$
$F'_i>1$ implies $b^i_j=k'_j/2$ for $j$ belonging to some set, call it
$I^i$. Then $(5.7a)$ becomes either
$$\eqalignno{
\sin(3\pi/k_i')\prod_{\ell\in I^i}\sin(\pi/k_\ell')\geq &{1\over F'_i}
\sin(\pi/k_i')\sp\quad {\rm if}\sp i\not\in I^i,&(5.7b)\cr
\sin(3\pi/k_i')\prod_{\ell\in I^i\atop \ell\ne i}\sin(\pi/k_\ell')\geq &
{1\over F'_i}\qquad\quad {\rm if}\sp i\in I^i.&(5.7c)\cr}$$
The hypotheses of this theorem were designed to ensure equations ($5.7b),
(5.7c$)
can never be satisfied. Thus $b^i$ will not be a fixed point of $\I_R$, so
$G'([b^i]'_j)=F'_i=1$.

Assume for now that $k_i\ne 3$. Then by $(5.3b$) there will be exactly three
$[a]$ such that the fusion rule
$N^L_{[f^i],[f^i],[a]}=1$. Thus there must be exactly three $[c]'_j$
such that $N^R_{[b^i]',[b^i]',[c]'_j}=1$. We claim this forces $b^i=J^i
f^{{\bar i}}$ for some simple current $J^i$, and some index ${\bar i}$.
Indeed,
if for example there was a $\ell$ for which $3<(b^i)_\ell<k_i'-3$, then there
would be at least four such $c$. The other possibilities are eliminated
similarly.

In fact we have $k_{{\bar i}}=k_{i}$. This follows from equation ($5.7a)$:
$$\sin(3\pi/k_i')\cdot \sin(\pi/k_{{\bar i}}')=\sin(\pi/k_i')\cdot\sin(3\pi/
k_{{\bar i}}'),\quad i.e.\sp\sin(\pi/k_i')=\pm\sin(\pi/k_{{\bar i}}').
\eqno(5.8a)$$

A similar conclusion holds if instead $k_i=3$. Then there must be exactly
two $[c]'_j$ such that $N^R_{[b^i]',[b^i]',[c]'_j}=1$.

This defines a function $\pi$ on $\{1,\ldots,r\}$, sending $i$ to ${\bar i}$
(define $\pi i
=i$ whenever $k_i=1$). We want to show $\pi$ is a conjugation, \ie that it is
also a bijection. If for some $i\ne j$ we have $\pi i=\pi j$, then
$$\sin(3\pi/k_i')\cdot\sin(3\pi/k_i')=\sin(9\pi/k_i')\cdot\sin(\pi/k_i'),
\eqno(5.8b)$$
\ie $\sin(\pi/k_i')=\pm\sin(3\pi/k_i')$, which only has $k'_i=4$ as a
solution. So $\pi$ defines a conjugation, and can be factored off. For
now on take $\pi=id$.

What we have shown is that for all $i$ with $k_i\ne 1$, $\tau([f^i])=
[J^if^i]'$ for some simple current $J^i$. Now take any $[a]_j$,
and write $[b]'_\ell=\tau([a]_j)$. Dividing $S^L_{[a]_j,[f^i]}=
S^R_{[b]'_\ell,[b^i]'}$
by $S^L_{[a]_j,[\rho]}=S^R_{[b]'_\ell,[\rho]'}$ gives us $b_i=a_i$ or
$k_i'-a_i$, in the usual way (see $(5.5d))$.
A similar conclusion occurs automatically whenever $k_i=1$.
Thus $M$ (or more precisely $M\cdot I^{\pi^{-1}}$) is a simple current
invariant, so by Thm.2 can be written in the form $(5.6$).\qquad QED
\medskip

The known $g^r_k$ ${\cal E}_{16}$-like exceptionals occur at levels where
at least 1 $k_i$ equals 16, or at least 2 equal 8, or at least two equal
4, or (see equations (5.12)) at least 8 equal 2.
Certainly there is no question that with a little more work the hypotheses
in Thm.4 can be weakened -- two results along these lines are given below.

\medskip\noindent{{\bf Theorem 5.}}\quad {\it Let $M$ be a physical
invariant. Suppose $M_{a\rho}\ne 0$ iff $a\in \I_L\rho$, and $M_{\rho b}\ne
0$ iff $b\in\I_R\rho$. Assume in addition that $\I_L$ and $\I_R$ have no
fixed points.
Then there exists a conjugation $\pi$ such that $MI^\pi$ is a simple
current invariant.}\medskip

\noindent{\it Proof.}\quad Look at the matrix $M'=MM^T$; as usual write it in
block form as $M'=\oplus B_i$, where each $B_i$ is indecomposable, and
where $B_{1}$ contains the $(\rho,\rho)$-entry. Then by Lemma 2(iii)
$B_1$ is the $\|\I_L\| \times \|\I_L\|$ matrix
$$B_1=\|\I_L\|\left(\matrix{1&\cdots&1\cr \vdots&&\vdots\cr 1&\cdots& 1\cr}
\right),\eqno(5.9a)$$
so we know the Perron-Frobenius eigenvalues all
satisfy $r(B_i)\le \|\I_L\|^2$.

Now choose any $a\in\p_L$, then there exists a $b\in\p_R$ such that
$M_{ab}\ne 0$. Suppose the $(a,a)$-entry of $M'$ is in $B_i$. Then $\forall
J,J'\in\I_L$,
$$(B_i)_{Ja,J'a}=(B_i)_{aa}=\sum_{c\in\p_R}M_{ac}^2\ge
\sum_{c\in\I_Rb}M_{ac}^2=\|\I_R\|M^2_{ab}.\eqno(5.9b)$$
Let $B^a_i$ denote the $\|\I_L\|\times\|\I_L\|$ matrix
$$(B_i^a)_{cd}=\left\{\matrix{\|\I_L\|M^2_{ab}&{\rm if}\sp c,d\in\I_La
\cr 0&{\rm otherwise}\cr}\right. .\eqno(5.9c)$$
Then $(B_i)_{cd}\ge (B^a_i)_{cd}\ge 0$ $\forall c,d$, so $r(B_i)\ge
r(B^a_i)$, with equality iff $B_i=B^a_i$. But $r(B_i^a)=\|\I_L\|^2
M^2_{ab}$. Therefore $B_i^a=B_i$, and $M_{ab}=1$.

What this means is that there is a bijection $\tau$ from $\p_L/\I_L$ onto
$\p_R/\I_R$, defined by $\tau[a]=[b]'$ iff $M_{ab}\ne 0$.
The equation $SM=MS$ says that, for any $a,a'\in\p_L$,
$$S_{aa'}=S_{\tau a,\tau a'}.\eqno(5.10a)$$
In order to understand this bijection $\tau$, we will mimic as closely
as possible the proof of Thm.1 in Sect.3. One complication is that here
the weights are constrained to come from $\p_L$ and $\p_R$.

Use $(5.3b$) to formally define the fusion rules $N^L_{[a][b][c]}$.
Now, the statement that $\I_L$ has no fixed points means that any $J\in
\I_L$ has $J_i=1$ for some $i$ with $k_i$ odd. This means that at most
one $d\in[c]$ will have $N_{abd}\ne 0$, so $N^L_{[a][b][c]}=0$ or 1, as in
Sect.3. Similarly for $N^R_{[a]'[b]'[c]'}$.

Choose $a=c$. We are interested in $[b]$ for some $b$ with all coefficients
odd, otherwise $N^L_{[a][a][b]}$ will vanish. Such $[b]$ will automatically
lie in $\p_L$. The number of such $[b]$ is precisely given by the LHS of
equation (3.8). Thus $(3.8)$ remains valid.

Consider first the odd $k_i$. Define $g^i=(1,\ldots,1,k_i'-2,1,\ldots,1)$.
Then $g^i\in \p_L$, and the argument of Thm.1 applies, and we find that
there is a conjugation $\pi_o$ acting only on the odd $k_i$ such that
$$(\pi_o\circ\tau(a))_i=a_i\sp {\rm or}\sp k_i'-a_i,\qquad{\rm for \sp
all}\sp i\sp{\rm with}\sp k_i\sp{\rm odd}.\eqno(5.10b)$$
It remains to find a conjugation $\pi_e$ acting only on the even $k_i$
for which the analogue of $(5.10b$) holds for even $k_i$; then $\pi=\pi_e
\circ\pi_o$ would be the desired conjugation, and $MI^\pi$ would be a
simple current invariant.

Consider any even $k_i$, then there exists a vector $e^i\in\p_L$ such that
$(e^i)_j=1+\delta_{ij}$ when $k_i$ is even (this again follows because
there are no fixed points). $(3.8$) applied to these, and using $(5.10b)$,
defines $\pi_e$, and the remainder of the argument follows as before.
\qquad QED\medskip

This theorem tells us \eg when all $k_i$ are odd, the only exceptional
physical invariants are ${\cal E}_{10}$-like. We will use it in Sect.7.
It should be possible to strengthen this result somewhat.

Note that if $M$ is strongly physical, with chiral algebras ${\cal C}_L=
{\cal C}(\I_L)$, ${\cal C}_R={\cal C}(\I_R)$, then $\I_L$ will have
fixed points iff $\I_R$ will. The reason is that if ${\cal C}(\I_L)$ has
no fixed points then it will have exactly $\alpha=\prod (k_i+1)/\|\I_L\|^2$
extended characters $ch_i$, while if it does have fixed points, then
it will have more than this number (even if some $G([a]_i)>1$). Because
$\|\I_L\|=\|\I_R\|$ (Lemma 2),
and because there must be a bijection between the extended characters,
the desired result follows.

Finally, we will now look at the simplest case with fixed points. The
motivation is to see if ${\cal E}_{16}$ belongs to an infinite series of
${\cal E}_{16}$-like exceptionals (it will also be used in Sect.7).

\medskip\noindent{{\bf Theorem 6.}}\quad {\it Let $M$ be a strongly physical
invariant, with no $k_i=2$ and at most one $k_i=4$. Suppose $M_{a\rho}\ne 0$
iff $a\in\{\rho,J\rho\}$, and $M_{\rho b}\ne 0$ iff $b\in\{\rho,J'\rho\}$.
Then there exists a permutation invariant $I^\sigma$ such that either
$M=M(J)\, I^\sigma$, or (rearranging the levels if necessary) $M=(
{\cal E}_{16}\otimes{\cal A}_{\hat{k}})\,I^\sigma$, where ${\cal A}_{\hat{k}}$
denotes the diagonal invariant of $\hat{k}=(k_2,k_3,\ldots,k_r)$.}\medskip

\noindent{{\it Proof.}}\quad The case where both $J$ and $J'$ have no fixed
points is considered in Thm.5, so we may suppose both $J$ and $J'$ have
fixed points. Without loss of generality put $J_i=1$ iff $i\le n$ for some
$n$. Then all $k_i$, for $i\le n$, will be even, and $\sum_{i\le n} k_i\equiv
0$ (mod 4). For any $i> n$ consider $(e^i)_j
=1+\delta^i_j$. The usual argument shows that it cannot get mapped by $\tau$
to a
fixed point, and in fact will get mapped to $[b^i]'=[J^ie^{{\bar i}}]'$ for
some ${\bar i}$ with $J'_{{\bar i}}=0$ and $k_{{\bar i}}=k_i$ ($J'_{{\bar i}}
=0$ follows because $J^ie^{{\bar i}}$ must be in $\p_R$). This defines
a conjugation, as usual, and factoring it off allows one to consider $J=J'$.
Then for any $a\in\p_L$, there exists a simple current $J^a$ such that
$\tau([a]_j)_i=[J^aa]_i$, for all $i>n$, and $1\le j\le f(a)$. In fact a
similar calculation
to that of equation (3.9) gives that there exist numbers $f_{ij}\in\{0,1\}$
such that
$$J^a_i\equiv \sum_{j=1}^r(a_j-1)f_{ji}\qquad({\rm mod}\sp 2),\qquad \forall
i>n.\eqno(5.11)$$

Consider first the case $n>1$. Define $(f^i)_j=1+2\delta^i_j$, for each
$i\le n$. It suffices to show $\tau([f^i])$ can never be a fixed point. But
this follows from (5.7): because $\|I^i\|\ge n>1$, and since there is
no $k_j=2$ and at most one $k_j=4$, we find that $(5.7c$) cannot
have a solution. Therefore $M$ will be a
simple current invariant, and we are done.

Now consider the case when $n=1$. By Thm.4 it suffices to consider $k_1=4
$, 8, 12 or 16. Again define $f^1=(3,1,1,\ldots,1)$.
Then $(\tau[f^1])_i=1$ for all $i>1$, by $(5.11)$. Then by $T$-invariance,
for $k_1=4$ both $[f^1]$ and $\tau[f^1]$ will be fixed points, and for
$k_1=8$ or 12 neither will be. So for $k_1\ne 16$, $M$
will be a simple current invariant, hence of the desired form.

The remaining $n=1$ case, with $k_1=16$, can be handled directly using (5.11),
or by using a somewhat simplified argument along the lines of the $k_1=16$
part of the proof of Thm.7, given below.\qquad QED\medskip

In other words, when $\|\I_L\|=2$ there really is only one
${\cal E}_{16}$-like exceptional, namely ${\cal E}_{16}$ itself.
Since there is another ${\cal E}_{16}$-like exceptional at level (8,8), it is
natural to guess
that they will also be found at $g^4_{4,4,4,4}$ and $g^8_{2,2,2,2,2,2,2,2}$
(Thm.2 tells us there will not be an exceptional at $g^{16}_{1,\ldots,1}$).
In fact, this is the case: let $\I_4$ be the set of all $r=4$ simple currents,
and let $\I_8^e$ denote those $r=8$ ones with an even number of nonzero
components, and let $\langle\c_a\rangle_4$ denote the sum $\sum_{b\in\I_4a}
\c_b^4$ (similarly for $\I_8^ea$), then
$$\eqalignno{Z(4^4)=&|\langle\c_{1111}\rangle_4|^2+
2[(\langle \c_{3111}\rangle_4+\langle\c_{1311}\rangle_4
+\langle\c_{1131}\rangle_4+\langle\c_{1113}\rangle_4)\c_{3333}^*+cc]&\cr
&2|\langle\c_{1133}\rangle_4+\langle\c_{3311}\rangle_4|^2+
2|\langle\c_{1313}\rangle_4+\langle\c_{3131}\rangle_4|^2+
2|\langle\c_{1331}\rangle_4+\langle\c_{3113}\rangle_4|^2&\cr&
+8(|\langle\c_{1333}\rangle_4|^2+|\langle\c_{3133}\rangle_4|^2
+|\langle\c_{3313}\rangle_4|^2+|\langle\c_{3331}\rangle_4|^2
+|\langle\c_{3333}\rangle_4|^2);&(5.12a)\cr
Z(2^8)=&|\langle\c_{11111111}\rangle_8^e|^2+8[\langle\c_{31111111}\rangle^e_8
\c_{22222222}^*+cc]+64|\c_{22222222}|^2.&(5.12b)\cr}$$
There is an analogue of $Z(2^8)$
for each $r=8s$: replace `8' in $(5.12b$) with $2^{4s-1}$ and 64 with
$2^{8s-2}$. For $s>1$ these are not fundamentally new, however, since they can
be obtained from $Z(2^8)$ by tensor products and multiplication by the
elementary simple current invariant associated with $\I_{8s}^e$.

Both $Z(4^4)$ and $Z(2^8)$ seem to be new. $Z(2^8)$ seems particularly
interesting, since it corresponds to a simple current chiral algebra
${\cal C}(\I^e_8)$, with $f(22222222)=2$ (not 128), and $G([22222222]_1)
=G([22222222]_2)=8$. The $S$-matrix for this extension is
$$S^e={1\over 2}\left(\matrix{1&1&1&1\cr 1&1&-1&-1\cr 1&-1&1&-1\cr
1&-1&-1&1\cr}\right).\eqno(5.13)$$
It is unitary and symmetric, and obeys the relations $S^{e2}=(S^eT^e)^3
=I$, as it should. The corresponding fusion rules can be computed from
Verlinde's formula, and are all found to be non-negative integers.
\bigskip\bigskip

\noindent{{\bf 6. An example: all $gcd(k_i',k_j')\le 3$}}\bigskip

Let us review what we have learned. For a given modular invariant $M$ of
$g^r_k$,
let ${\cal K}^R_M(a)$ be the set of all $b\in P^r_k$ such that
$M_{ab}\neq 0$; define $\R^L_M(b)$ to be those $a\in P^r_k$ for
which $M_{ab}\ne 0$. We know (Thm.3 of \COMM{} together with equation
$(4.2b)$ above) that for a given physical invariant $M$, $\R^L_M(\rho_r)=
\{\rho_r\}$ iff
$\R^R_M(\rho_r)=\{\rho_r\}$ iff $M$ is a permutation invariant, in which
case it is listed in (3.5). If for all $a$, both $\R^L_M(a),\R^R_M(a)
\subset\I a$, then $M$ is a simple current invariant, and will appear in
Thm.2.

We are interested in finding all physical invariants for a given $k$. Let
$\R^r_k(a$) denote the $a$-couplings of $g^r_k$, \ie the union over all
$g^r_k$ physical invariants $M$ of the set $\R^L_M(a)$, or equivalently
of the set $\R^R_M(a)$.
For almost every $k$, experience tells us $\R^r_k(\rho_r)\subset \I\rho_r$,
in which case Thm.4 applies; Thm.3 says $\R^r_k(\rho_r)\not\subset\I\rho_r$
iff there
is an ${\cal E}_{10}$-like exceptional of $g^r_k$. So our strategy is to
try to find the $a$-couplings $\R^r_k(a)$, and in particular the
$\rho_r$-couplings $\R^r_k(\rho_r)$.

There are three main constraints. Let $b\in {\cal K}^r_k(a)$.
Then by $T$-invariance it must satisfy
$$\sum_{i=1}^r {a_i^2\over k_i'}\equiv  \sum_{i=1}^r{b_i^2\over k_i'} \sp
({\rm mod\sp} 4).\eqno(6.1a)$$
Also, the relation $MS=SM$ and the fact that $S^{(k)}_{\rho c}>0$ for all
$c$ implies for any positive invariant $M$ that
$$s(b)\equi \sum_{c\in P^r_k} M_{\rho c} S^{(k)}_{cb}\ge 0,\qquad \forall
b\in P^r_k,\eqno(6.1b)$$
with $s(b)=0$ iff $M_{cb}=0$ $\forall c\in P^r_k$.

The third tool is the {\it parity rule} \Paru. To formulate it, we
must make the following definitions.

Choose any $a_i\in \Z$, $i=1,\ldots,r$. For any $x,y$, by $\{x\}_y$ we
will mean the unique number $\equiv x$ (mod $y$) lying between 0 and $y$.
For each $i$ define $a_i^+$ and $\epsilon_i$ by:
$a_i^+=\{a_i\}_{2k_i'}$ and $\epsilon_i=+1$ if $0<\{a_i\}_{2k_i'}<k_i'$;
$a_i^+=2k_i'-\{a_i\}_{2k_i'}$ and $\epsilon_i=-1$ if
$k_i'<\{a_i\}_{2k_i'}<2k_i'$;
$a_i^+=0$ and $\epsilon_i=0$ if $\{a_i\}_{2k_i'}$ equals either
0 or $k_i'$. We call $\epsilon(a)=\epsilon_1\cdots\epsilon_r$ the {\it
parity} of $a$. Note that $\epsilon(a)\in\{\pm 1,0\}$, and when it
is nonzero, $a^+\in P_k^r$.

Now choose any $a,b\in P_k^r$. Let $L_k$ be the set of all integers
$\ell$ coprime to $2k_1'\cdots k_r'$. Then $\epsilon(\ell a)
\epsilon(\ell b)\ne 0$. More importantly \COMM{}
for any level $k$ modular invariant $M$, and any $\ell\in L_k$,
$$M_{ab}=\epsilon(\ell a)\,\epsilon(\ell b)\,M_{(\ell a)^+,
(\ell b)^+}.\eqno(6.2a)$$
Thus $\R_k^r((\ell a)^+)=(\ell\R_k^r(a))^+$, for all $\ell\in L_k$,
$a\in P^r_k$.
A similar statement to $(6.2a$) holds for any RCFT \CG, and plays an
important role in the ${\hat A}_2$ classification \GR, the computer
search in \HK, and the heterotic classification in \GH. We call it the
parity rule.

Its main value for our purpose lies in its immediate consequence:
$$b\in\R_k^r(a)\Rightarrow \eps(\ell a)=\eps(\ell b)\qquad\forall \ell
\in L_k.\eqno(6.2b)$$

Together, $(6.1)$ and $(6.2b$) constitute extremely strong constraints
on the sets $\R_k^r(\la)$. No assumptions beyond (P1)-(P3) were needed to
derive them.

We will first look at $(6.2b)$ for the case $r=1$.

\medskip\noindent{{\bf Lemma 3.}}\quad (a) {\it Let $n$ be
odd, and choose any integer
$0<a<n$. Suppose $0<b<n$ satisfies, for all $\ell$ coprime to $2n$,
the relation}
$$\{\ell a\}_{2n}<n\quad{\rm iff}\quad \{\ell b\}_{2n}<n.\eqno(6.3)$$
{\it Then either $b=a$ or $b=n-a$.}

\noindent{(b)} {\it Let $n$ be even. Then for $a=1$, the solutions $b$ to
(6.3) are:}
\item{} {\it for} $n\ne 6,10,12,30$: $b=1$ {\it and} $n-1$;

\item{} {\it for} $n=6$: $b=1,3,5$;

\item{} {\it for} $n=10$: $b=1,3,7,9$;

\item{} {\it for} $n=12$: $b=1,5,7,11$; {\it and}

\item{} {\it for} $n=30$: $b=1,11,19,29$.

\medskip\noindent{\it Proof.}\quad (a) For $n$ odd, each $\ell_i=2^i+n$ is
coprime
to $2n$. Write the binary expansions $x=a/n=\sum_{i=1}^\infty x_i 2^{-i}$,
$y=b/n=\sum_{i=1}^\infty y_i 2^{-i}$, where each $x_i,y_i\in\{0,1\}$.
Then for $i=1,2,\ldots$, taking $\ell=\ell_i$ in (6.3) gives us $a+x_i\equiv
b+y_i$ (mod 2). If $a\equiv b$ (mod 2), this forces $a=b$; if $a\not\equiv
b$ (mod 2) this forces $a=n-b$.

(b) For $n$ even and $a=1$, all the work was done in Claim 1 of Sect.4 of
\GR. There
we found all {\it odd} $b$ solutions to (6.3). That any solution $b$ to (6.3)
must be odd, when $n$ is even and $a=1$, follows by taking $\ell=n-1$ there.
\qquad QED\medskip

Remember in using Lemma 3 that $n$ there plays the role of height $k+2$, not
level $k$.
If we define an {\it anomolous coupling} here to be solutions $a,b$ to (6.3),
where both $a\ne b$ and $a+b\ne n$, then Lemma 3(a) says that for $n$ odd
there are no anomolous couplings.  However, for $n$ even they are common.
The list of all anomolous couplings for $n\le 30$ are:
$$\eqalignno{n=6:&\sp\{1,3,5\};&\cr n=10:&\sp \{1,3,7,9\};&\cr n=12:&
\sp\{1,5,7,11\},\sp\{2,6,10\};&\cr
n=18:&\sp\{3,9,15\};&(6.4)\cr n=20:&\sp\{2,6,14,18\};&\cr n=24:&\sp
\{2,10,14,22\},\sp\{4,12,20\};&\cr
n=30:&\sp\{1,11,19,29\},\sp\{3,9,21,27\},\sp\{5,15,25\},\sp\{7,13,17,23\}.
&\cr}$$

Equation (6.4) should be read as follows. If $a\ne b$ is a solution to (6.3),
and $a+b\ne n$ then $a$ and $b$ will both belong to one of the sets listed in
(6.4). Conversely, two elements $a\ne b$ of a common set in (6.4), which
do not satisfy $a+b=n$, will be an anomolous coupling for that height $n$.

When $r=1$ it is convenient to identify a weight $a$ with its Dynkin label
$a_1$. Also, when the level $k$ is understood we will write $\bar a=k'-a$.

One useful consequence of Lemma 3 is that it will permit us to find all
the {\it positive} invariants of $A_{1,k}$. Recall that a modular invariant
is called positive if all its coefficients $M_{\la\mu}\ge 0$ (but they need
not be integers). Equations (6.1),(6.2) hold for any $b\in\R^R_M(a)$, for any
positive invariant $M$. The
main reason the positive invariants of $A_{1,k}$ will be useful is through
projecting \BOU{} out $r-1$ of the $A_1$ factors. See \CL{} for an example.

\medskip\noindent{{\bf Lemma 4.}}\quad {\it Any positive invariant of $A_{1,k}$
can be written as a linear combination (with positive coefficients) of
the 1, 2, or 3 physical invariants at that level.}\medskip

\noindent{{\it Proof.}}\quad The three exceptional levels can be done on a
computer, or by hand using (6.4). For $k$ odd, the result follows from
Lemma 3(a) and $T$-invariance, and the fact that the only modular invariants
which are
diagonal matrices will be, for any algebra and level, a scalar multiple of
the identity. Therefore we need to consider only even $k\ne 10,16,28$.

First consider $k\equiv 2$ (mod 4). By Lemma 3(b) and $T$-invariance,
$\R^R_M(1)=\R^L_M(1)=\{1\}$ for any positive invariant $M$.

Write $\R_M(a)=\R^R_M(a)\cup\R^L_M(a)$. Suppose we know
$\R_M(a)\subset\{a,\bar a\}$ for all $a$. Then the $(1,a)$, $(a,1)$,
$(1,\bar a)$ and $(\bar a,1)$ entries of $SM=MS$ tell us $M_{11}=
M_{aa}$ for $a$ odd, and $M_{11}=M_{aa}+M_{a\bar a}$, $M_{aa}=
M_{\bar a \bar a}$,
$M_{a\bar a}=M_{\bar a a}$ for $a$ even. Also $(SM)_{2a}=(MS)_{2a}$ tells us
$M_{22}-M_{2\bar 2}=M_{aa}-M_{a\bar a}$, for $a$ even. Then $M=M_{22}{\cal A}_k
+M_{2\bar 2}{\cal D}_k$, and we would be done. So we want to show $\R_N(a)
\subset \{a,\bar a\}$.

By an argument similar to that used in deriving $(5.5c$), by looking at
$N^\infty=
{\rm lim}_{n\rightarrow\infty}\{(M+M^T)/(2M_{11})\}^n$ we get the following
expression:
$$C_a\sum_{c\in W(a)}S_{1a}S_{1c}S_{cb}=C_b\sum_{d\in W(b)}S_{ad}
S_{1d}S_{1b}\quad {\rm where}\sp C_a=\sum_{c\in W(a)}S_{1c}^2\eqno(6.5a)$$
for all $a,b$, where $W(a)\supset\R_M(a)$ is some set of weights (namely, the
weights in the indecomposable block of $N^\infty$ containing $a$). Suppose
we know $W(a)=\{a\}$ for some $a$, then for any $b$
$${S_{ab}\over S_{1b}}={S_{ad}\over S_{1d}}\qquad\forall d\in W(b).
\eqno(6.5b)$$
If instead we know $W(a)=\{a,\bar a\}$ for some $a$, then $(6.5b)$ will hold
for any odd $b$.

If in equation (6.5$b$) we have $a=3$, then $d=b$ or $\bar b$ and we are done,
so it suffices here to show $W(3)=\{3\}$. $W(a)=\{a\}$ does hold for any $a$
coprime
to $2k'$, by $(6.2a$) and the fact that $W(1)=\{1\}$. If 3 does not divide
$k'$, we are done. Otherwise $(6.5b$) gives us $S_{3a}/S_{13}=S_{ad}/S_{1d}$
for any $a$ coprime to $2k'$ and $d\in W(3)$. But the LHS will always be
$\ge 1$ in absolute value, which forces either $d$ or $\bar d$ to divide $k'$.
$T$-invariance now forces $d=3$ or $\bar 3$, and we are done.

The proof for $k\equiv 0$ (mod 4) is similar. Here we get $\R_M(3)\subset
W(3)\subset\{3,\bar 3\}$, provided we avoid the exceptional case $k=16$. By
$(6.5b$) this forces $\R_M(a)\subset\{a,\bar a\}$ for all odd $a$. From the
usual $MS=SM$ arguments, we get $M_{aa}=M_{11}\ge M_{1\bar 1}=M_{a\bar a}$ for
all odd $a$. Replace $M$ with $M'=M-M_{1\bar 1}{\cal D}_k$. $M'$ will be
positive,
it will have only 1 as a $\rho$-coupling, and $a=3$ in $(6.5b$) together
with $MS=SM$ forces $M'=M_{22}{\cal A}_k$.\qquad QED\medskip

Consider now the case where each $k_i'$ is coprime
to each $k_j'$, $i\ne j$.
Given any $\ell\in L_k$, choose any  $\ell_i\equiv \ell$ (mod $2k_i')$.
Then each $\ell_i$ is coprime to $2k_i'$. The converse is also true: given
any set
$(\ell_1,\ldots,\ell_r)$, where each $\ell_i$ is coprime to $2k_i'$,
there is an
$\ell\in L_k$ such that $\ell_i\equiv \ell$ (mod $2k_i')$ (this follows
from the Chinese Remainder Theorem).

Fix $i$. For each $\ell_i$ coprime to $2k_i'$, let $\ell\in L_k$ correspond to
the $r$-tuple $(1,\ldots,\ell_i,\ldots,1)$. Then $(6.2b$) collapses to the
one-dimensional
$$\eps(\ell_ia_i)=\eps(\ell_ib_i), \qquad\forall \ell_i\sp{\rm coprime\sp
to\sp}2k_i',\eqno(6.6)$$
which was considered in Lemma 3. A little more care shows that (6.6) still
holds, if all we have here is each $gcd(k_i',k_j')\leq 3$. Indeed, for a
given $\ell_i$ coprime to $2k_i'$, choose $\ell_j=\pm 1$, $j\ne i$, according
to the following rule: if $g_{ij}=gcd(k_i',k_j')=1$, choose $\ell_j=+1$;
otherwise choose $\ell_j\equiv \ell_i$ (mod 2$g_{ij}$). Then, again by the
Chinese Remainder Thm., $\exists\ell\in L_k$ such that $\ell\equiv\ell_j$
(mod $2k_j'$), $\forall j$. Since $\eps_{k_j}(\pm a_j)=\pm\eps_{k_j}(a_j)$,
for any $a_j$, $(6.2b)$ reduces to (6.6) for this $\ell$. We now know
enough for a quick proof of the following result; the only hard part as usual
is handling the 3 exceptional levels.

\medskip\noindent{{\bf Theorem 7.}}\quad {\it Suppose each gcd$(k_i',k_j')\leq
3$, for $i\ne j$. Then the only strongly physical invariants belonging to these
levels $k$ are the simple current invariants, classified in Thm.2, together
with the products (reordering the levels if necessary) $({\cal E}_{10}\otimes
{\cal A}_{\hat{k}})\,{\cal D}_k$, ${\cal E}_{16}\otimes {\cal D}_{\hat{k}}$,
and ${\cal E}_{28}\otimes {\cal D}_{\hat{k}}$, where ${\cal A}_k$ denotes
the diagonal invariant of level $k$, ${\cal D}_k$ denotes any simple current
invariant of level $k$, and $\hat{k}=(k_2,\ldots,k_r)$.}
\medskip

\noindent{{\it Proof.}}\quad We would like to use Lemma 3 and (6.6) to get
that $\R^r_k(\rho)\subset \I\rho$, and then Thm.4 to get the desired result.
But the first step requires that we avoid $k_i$ equal to 4, 8, 10 or 28, and
the second step that we avoid 2, 4, 8, 12 and 16.

Choose any $a$, and any $b\in \R^R_k(a)$.
Consider first that $k_i=2$ or 12. Because it (\ie $n=4,14$) is not on the
list of (6.4), (6.6) implies $b_i=a_i$ or $\bar a_i$. If $k_i=8$, then (6.4)
and $T$-invariance (in particular $(6.1a)$ multiplied by $\prod_j k_j'/5$,
taken mod 1) also implies $b_i=a_i$ or $\bar a_i$.

Next suppose $k_i=4$. Then the gcd condition says no $k_j=10,16,28$, so for
all $j\ne i$, $a_j=1$ implies $b_j=1$ or $\bar 1$. Then by $T$-invariance
$\R^r_k(\rho)\subset \I\rho$. The familiar argument shows that $f^i=(1,
\ldots,3,\ldots,1)$,
can only couple to $Jf^i$ for some simple current $J$. As usual
this forces $b_i=a_i$ or $\bar a_i$.

Thus as long as no $k_i=10$, 16, 28, the proof of Thm.4 will carry through,
and the strongly physical invariant will be a simple current invariant, listed
in Thm.2.
So it suffices to consider $k_1$, say, equal to one of 10, 16, 28. The gcd
condition then says that no other $k_i=4,10,16,28$, so
$$b_i=a_i\sp{\rm or}\sp \bar a_i,\sp \forall i>1.\eqno(6.7a)$$

Consider first $k_1=16$. Then $\R^r_k(\rho)\subset\I\rho$, so by Thm.3
we have an automorphism $\tau$ between simple current chiral algebras
${\cal C}(\I_L)$, ${\cal C}(\I_R)$. Let $M$ be any strongly physical
invariant, and write $f^1=(3,1,\ldots,1)$. If
$\tau[f^1]$ is not a fixed point, then $M$ is a simple current invariant
by the usual argument. So suppose $\tau[f^1]$ is a fixed point, \ie
the first component $(\tau[f^1])_1=9$, and $J^1=(1,0,\ldots,0)\in\I_L\cap
\I_R$. Write $\hat{a}
=(a_2,\ldots,a_r)$, and $\hat{\I}_R$ be that subset of $\I_R$ which
fixes the first component. So $\I_R$ is
spanned by $J^1$ and $\hat{\I}_R$.
Let $\hat{\p}_L$ be the set of all $\hat{a}\in
\p^{r-1}_{\hat{k}}$ such that there exists an $x<18$, $b\in\p_R$, such that
$M_{x\hat{a},b}\ne 0$. First note that for any $a$, (6.4) tells
us $(\tau[a])_1=a_1$ or $\bar a_1$, if $a_1=1$, 5 or 7. Choose any
$\hat{a}\in\hat{\p}_L$, and suppose $\tau([1\hat{a}]_1)=[1\hat J\hat{a}]'_i$.
Choose any $[x\hat{a}]_j$ and any $\hat{c}\in\hat{\p}_L$ with components
${\hat c}_\ell=1$ or 2 (so ${\hat F}_{\hat c}=1)$,
and write $\tau([x\hat{a}]_j)=[y{\hat J}'\hat{a}]'_\ell$ and $\tau[1\hat{c}
]=[1\hat{c}']'$. Comparing
$$\eqalignno{{G([1\hat a]_1)\over \hat{F}_{\hat{a}}}S_{1\hat{a},1\hat{c}}=
{G'([1\hat J\hat a]_i)\over\hat{F}_{\hat J\hat{a}}}'
S_{1\hat J\hat{a},1\hat{c}'},&\quad i.e.\sp
\hat{S}_{\hat{a}\hat{c}}=\hat{S}_{\hat{a}\hat{c}'}\cdot
-1^{\hat J\cdot(\hat{c}'-\hat{\rho})}\alpha&(6.7b)\cr
{G([x\hat a]_j)\over F^1_x\hat{F}_{\hat{a}}}S_{x\hat{a},1\hat{c}}=
{G'([y\hat J'\hat a]_\ell)\over F^1_y\hat{F}_{\hat J'\hat{a}}'}
S_{y\hat J'\hat{a},1\hat{c}'},&\quad i.e.\sp
\hat{S}_{\hat{a}\hat{c}}=\hat{S}_{\hat{a}\hat{c}'}\cdot
-1^{\hat J'\cdot(\hat{c}'-\hat{\rho})}\beta,&(6.7c)\cr}$$
where $\hat S$ is the $S$-matrix for $g^{r-1}_{\hat k}$, and $\alpha,\beta$
are the obvious combinations of $G,G',\hat F,\hat F'$. By the usual argument
(see (4.9)) we see that we can choose $\hat J$ and $\hat J'$ so that $(\hat J
\hat J')\cdot(\hat{c}'-\hat{\rho})\equiv 0$ (mod 2). But any
$\hat{d}\in\hat{\p}_L$ is congruent (mod 2) to some $\hat{c}\in\hat{\p}_L$
whose components are 1's and 2's.
So we get from the (4.3) argument that $\hat J\hat J'\in\hat{\I}_R$. In other
words,
$$\tau([1\hat{a}]_1)=[1\hat{b}]'_i\sp\Rightarrow\sp \forall x,j,\sp \exists
y,\ell \sp{\rm such\sp that}\sp\tau([x\hat{a}]_j)=[y\hat{b}]'_\ell.
\eqno(6.7d)$$

Now define the projection $M'$ of $M$ by ${\cal A}_{16}$:
$$M'_{\hat{a}\hat{b}}=\sum_{x=1}^{17}M_{x\hat{a},x\hat{b}}.\eqno(6.8a)$$
Then $M'$ is a positive invariant. $(6.7a$) tells us that $M'$ is a
(non-physical) simple current invariant.
Note that $M'_{\hat{\rho}\hat{\rho}}=7$. In fact $(6.7d)$ implies
$M'_{\hat J\hat{\rho},\hat{\rho}}=0$ unless $\hat J\in{\hat\I}_L$,
in which case it equals 7; similarly, $M'_{\hat \rho,\hat J'}=0$ unless $\hat
J'\in{\hat \I}_R$, in which case it also equals 7.
Then (4.3) says $\hat{\I}_L\cdot(\hat{\p}_L-\hat{\rho})\equiv
0$ (mod 2), so by the (4.4) calculation we get that
$$\sum_{\hat{\mu}}M'_{\hat{\la}\hat{\mu}}=7\|\hat{\I}_L\|.\eqno(6.8b)$$
{}From $(6.8b$) we see $(\tau[3\hat{a}]_i)_1=9$. Gathering all the results
we have, we get that $M={\cal E}_{16}\otimes{\cal D}_{\hat{k}}$ for the
(physical) simple current invariant ${\cal D}_{\hat{k}}={1\over 7}M'$.

Next consider $k_1=28$. Again, $(6.7a$) is satisfied. Without loss of
generality suppose $\R^R_M(\rho)\not\subset\I\rho$.\smallskip

\noindent{{\bf Claim 2.}}\quad {\it Suppose $M_{\rho,11\hat{b}}\ne 0$ or
$M_{\rho,19\hat{b}}\ne 0$. Then }
$$\R^L_M(\rho)=\hat{\I}_L\{(1\hat{\rho}),\,(11\hat{\rho}),\,(19\hat{\rho}),\,
(29\hat{\rho})\},\sp\R^R_M(\rho)=\hat{\I}_R\{(1\hat{\rho}),\,
(11\hat{\rho}),\,(19\hat{\rho}),\,(29\hat{\rho})\},$$
{\it for some groups $\hat{\I}_L,\hat{\I}_R$ of simple currents fixing
the first component.}\smallskip

\noindent{{\it Proof of claim.}}\quad The proof is based on a somewhat
tedious application of $(6.1b$). We will sketch some of the details.
Define $\hat{\I}_R$ to be those $\hat J$ such that $M_{\rho,1\hat J\rho}\neq
0$. It will be a group, by Lemma 2. Let $\|\hat{\I}_R\|n_{x}=\sum_{\hat{b}}
M_{\rho,x\hat{b}}$ for $x=11,19,29$. These $n_x$ will be
integers by Lemma 2. Putting $a=(3\hat{\rho})$ in $(6.1b)$ gives us
the relation $1-n_{11}-n_{19}+n_{29}\ge 0$. Continuing in this way we
force $n_{11}=n_{19}=n_{29}=1$. Then there exists simple currents ${\hat J}^x$
such that $M_{\rho,x\hat{b}}\ne 0$ iff $\hat{b}\in\hat{\I}_R{\hat J}^x
\hat{\rho}$, in which case $M_{\rho,x\hat{b}}=1$. To show ${\hat J}^x\in
\hat{\I}_R$ for each $x$, it suffices to show that for each $\hat a$
satisfying $\hat \I_R
\cdot(\hat a-\hat \rho)\equiv 0$ (mod 2), then also ${\hat J}^x
\cdot(\hat a-\hat \rho)\equiv 0$. This follows from $(6.1b$) -- \eg if
for some such $\hat a$ we had ${\hat J}^{x}\cdot(\hat a-\hat \rho)\equiv
1$ (mod 2) for all $x=11,19,29$, then take $a=(1,\hat a)$ in $(6.1b$).

That $\R^L_M(\rho)$ must also contain anomolous $\rho$-couplings follows
from $(SM)_{\rho\rho}=(MS)_{\rho\rho}$.\qquad QED to claim

$(6.1b$) and $T$-invariance (using $(4.7a$)) now tell us that
$M_{x\hat{a},y\hat{b}}\neq 0$ implies either $x,y\in\{1,11,19,29\}$
or $x,y\in\{7,13,17,23\}$.

Define the projection
$$M'_{\hat{a}\hat{b}}={1\over 8}\sum_{x=1}M_{x\hat{a},x\hat{b}}=
M_{1\hat{a},1\hat{b}}; \eqno(6.9a)$$
the second equality follows from $(6.2a)$. Obviously $M'$ is a
(physical) simple current invariant. Project again, giving
$$M''_{xy}={1\over\|\hat{\I}_R\|^2}\sum_{\hat{a},\hat{b}}
M_{x\hat{a},y\hat{b}}M'_{\hat{a},\hat{b}}.\eqno(6.9b)$$
Lemma 4 and the fact that $\R^R_{M''}(3)=\{\}$ implies $M''={\cal E}_{28}$.
This implies that $M={\cal E}_{28}\otimes M'$.

Finally, consider $k_1=10$. Again, $(6.7a$) is satisfied. A similar argument
to the one used in the proof of Claim 2 gives us:

\smallskip\noindent{{\bf Claim 3.}}\quad Suppose $\R^R_M(\rho)\not\subset
\I\rho$. Then either

\item{case 1:}\qquad $\R^R_M(\rho)=\hat{\I}_R\{(1\hat{\rho}),\,(7\hat{\rho})\}
\sp{\rm and}\sp \R^L_M(\rho)=\hat{\I}_L\{(1\hat{\rho}),\,(7\hat{\rho})\}$;

\item{case 2:}\qquad $\R^R_M(\rho)=\hat{\I}_R\{(1\hat{\rho}),\,(7\hat{\rho}),
\,(5{\hat J}^R\hat{\rho}),\,(11{\hat J}^R\hat{\rho})\}$

\item{} \qquad\qquad and $\R^L_M(\rho)=\hat{\I}_L\{(1\hat{\rho}),\,
(7\hat{\rho}),\,(5{\hat J}^L\hat{\rho}),\,(11{\hat J}^L\hat{\rho})\}$;

\noindent{where} $\hat{\I}_L,\hat{\I}_R$ are groups of simple currents
fixing the first component and ${\hat J}^L,{\hat J}^R$
are simple currents.\smallskip

First look at case 1. $(6.2a$) tells us that \eg $M_{1\hat{a},1\hat{b}}=
M_{x\hat{a},x\hat{b}}$ for $x=5,7,11$. Also, the equation $M=SMS$ produces
expressions like
$$\eqalignno{M_{1\hat{a},{c}}=&\sqrt{{1\over 6}}\bigl\{ \sin(\pi/12)\,
(A^1_{\hat{a}c}+A^{11}_{\hat{a}c})+\sin(5\pi/12)\,(A^5_{\hat{a}c}+
A^7_{\hat{a}c})+\sin(\pi/3)\,(A^4_{\hat{a}c}+A^8_{\hat{a}c})\bigr\},&\cr
{\rm where}&\quad A^{x}_{\hat{a}c}=\sum_{\hat{b},d}
\hat{S}_{\hat{a},\hat{b}}M_{x\hat{b},d}S_{dc}.&(6.10)\cr}$$
Comparing the expressions for $M_{x\hat{a},c}$ for $x=1,5,7,11$, and noting
from $T$-invariance that at least two of these vanish, we get that
$M_{1\hat{a},c}=M_{7\hat{a},c}$ and $M_{5\hat{a},c}=M_{11\hat{a},c}$.
Familiar arguments tell us that all row and column sums must be equal to
$2\|{\hat \I}_L\|$. It can be shown, by evaluating $MS=SM$ at $(1\hat{a}, 4
\hat{\rho})$, that there is no triple $\hat{a},\hat{b},\hat{c}$ such that
both $M_{1\hat{a},1\hat{b}}\neq 0$ and $M_{1\hat{a},5\hat{c}}\ne 0$.
Suppose there is a $\hat{a},\hat{c}$ such that $M_{1\hat{a},5\hat{c}}\ne 0$.
The same calculation tells us there exists a simple current $\hat J$ such
that $M_{4\hat{J}\hat{\rho},4\hat{\rho}}=1$, and it satisfies ${\hat J}\cdot
(\hat{a}-\hat{\rho})\equiv 1$ (mod 2). By $T$-invariance, the simple current
$J=(1,\hat J)$ has norm $J^2\equiv 2$ (mod 4); write $\tilde M=M\,I^J$. Then
${\tilde M}_{1\hat{b},5\hat{d}}=0$ for all $\hat{b},\hat{d}$. It is now
easy to show $\tilde M={\cal E}_{10}\otimes {\tilde M}'$, where
$${\tilde M}'_{\hat{a}\hat{b}}={1\over 6}\sum_{x=1}^{11}
{\tilde M}_{x\hat{a},x\hat{b}}={2\over 3}
{\tilde M}_{1\hat{a},1\hat{b}}+{1\over 3}
{\tilde M}_{4\hat{a},4\hat{b}}.\eqno(6.11)$$

All that remains is case 2. Then $J=(1,{\hat J}^R)\in\I_R$, so by Lemma 2
$M_{1\hat{a},1\hat{b}}=M_{1\hat{a},11{\hat J}^R\hat{b}}$, etc. The proof
that $M=({\cal E}_{10}\otimes {\cal A}_{\hat{k}})\,M(J)\,({\cal A}_{10}\otimes
{\cal D}_{\hat{k}})$ for some simple current invariant ${\cal D}_{\hat{k}}$
now proceeds as in (6.10) above. \qquad QED\medskip

Throughout the proof of Thm.7, we used the easily verified fact that
if $M=M'\otimes M''$ and $M$ and $M'$ are both modular invariants, then
so will be $M''$.

What is so special about the numbers $\ell\le 3$ is that they are the only
ones with the following property: any $n$ coprime to $2\ell$ satisfies
$n\equiv \pm 1$ (mod $2\ell$). Nevertheless,
Thm.7 could be strengthened with more effort, but as it stands it provides
an easy example of the power of equations (6.1),(6.2). Those equations will
also be used in the following section. Note that Thm.5 tells us that
when all $k_i$ are odd, Thm.7 will apply to any (not necessarily strongly)
physical invariant. For $r=2$, it is also possible to show that Thm.7 holds
for any physical invariant.

\bigskip\bigskip
\noindent{{\bf 7. The su(2)$\oplus$su(2) classification}}\bigskip

In this section we apply all that we have learned in the previous sections,
to find all strongly physical invariants for $A_1\oplus A_1$ at all levels
$k=(k_1,k_2)$. It suffices to find all ${\cal E}_{10}$-like and ${\cal
E}_{16}$-like exceptionals; the regular series (\ie the simple current
invariants and their conjugations) are listed in the Appendix of \GH.
First we will find all ${\cal E}_{16}$-like exceptionals, as well as find a
characterization of all of the obvious (\ie non-sporatic) ${\cal E}_{10}$-like
exceptionals.
The previous sections suggest to look at the $\rho$-couplings $\R_M=\R^L_M
(\rho)\cup\R^R_M(\rho)$.

Throughout this section, $M$ will denote a strongly physical invariant of
$g^2_k$.
Let ${\cal A}_k$, ${\cal D}_k$, ${\cal E}_k$ denote the $A_{1,k}$ physical
invariants. Write $J^1=(1,0)$, $J^2=(0,1)$, and $J^{12}=(1,1)$.
A valuable result is that for all $21\ge k_1\ge k_2$, along with $k_1=28$ and
$21\ge k_2$, all physical invariants have been found using explicit
calculations
using the lattice method \Cmtr. Small levels often appear below as special
cases, and this permits us to dismiss them.

\medskip\noindent{{\bf Proposition.}}\quad (a) {\it Suppose  ${\cal K}_M
\subset \I\rho=\{1,k_1-1\}\times\{1,k_2-1\}$.
Then $M$ will lie on one of the regular series, or will be one of the
exceptionals
${\cal E}_{16}\otimes{\cal A}_{k_2}$, ${\cal E}_{16}\otimes{\cal D}_{k_2}$,
${\cal E}_{16}\otimes{\cal E}_{16}$, ${\cal A}_{k_1}\otimes{\cal E}_{16}$,
${\cal D}_{k_1}\otimes{\cal E}_{16}$ (or their conjugations, if $k_1=k_2=16$),
or the $k=(4,4)$ or $(8,8)$ exceptionals.}

\item{(b)} {\it Suppose for $k=(28,k_2$), we have $\R_M\subset
\{1,11,19,29\}\times\{1,k_2'-1\}$. Then either $M$ will be listed in} (a),
{\it or
$M={\cal E}_{28}\otimes{\cal A}_{k_2}$
or (if $k_2$ is even) ${\cal E}_{28}\otimes{\cal D}_{k_2}$ or (if $k_2=16$)
${\cal E}_{28}\otimes{\cal E}_{16}$, or (if $k_2=3$) the $k=(28,3)$
exceptional.}

\item{(c)} {\it Suppose for $k=(10,k_2$), we have $\R_M\subset\{1,5,7,11\}
\times\{1,k_2-1\}$. Then either $M$ will be listed in} (a), {\it or $M=
{\cal E}_{10}\otimes {\cal A}_{k_2}$, or (if $k_2$ is even) ${\cal E}_{10}
\otimes {\cal D}_{k_2}$ or $({\cal E}_{10}\otimes {\cal A}_{k_2})\,M(J^{12})$,
or (if $k_2=16$) ${\cal E}_{10}\otimes{\cal E}_{16}$, or (if $k_2=2$) the
$k=(10,2)$ exceptional}. \medskip

\noindent{{\it Proof.}}\quad (a) Thms.4 and 6 of Sect.5 prove this for
almost all levels. The remaining cases are:

\item{(i)} $k=(2,k_2)$, $k_2\equiv 2$ (mod 4), $\I_L=\I_R=\{0,J^{12}\}$;

\item{(ii)} $k=(2,k_2)$, $k_2\equiv 0$ (mod 4), $\I_L=\I_R=\{0,J^2\}$;

\item{(iii)} $k=(k_1,k_2)$, $k_1\equiv k_2\equiv 0$ (mod 4), $\I_L=\I_R=
\{0,J^1,J^2,J^{12}\}$, and $k_1\le 16$;

\item{(iv)} $k=(4,4)$, $\I_L,\I_R\ne\{0\}$.

(iv) has already been worked out explicitly \HK. In all other cases, Thm.3
applies, and we have a bijection $\tau$ between simple current extensions.

(i): \quad First note that $\tau$ will not send $f^2=(1,3)$ to a fixed
 point,
except possibly for $k=(2,10)$ (the usual $(5.7a)$ argument works here). The
familiar fusion rule argument then says $\tau[f^2]=[Jf^2]$ for some
simple current $J$, and $T$-invariance then forces $J\in\I_L$. Then for any
$\la\in\p_L$, the $(5.5d$) argument says $(\tau[a])_2=a_2$ or ${\bar a}_2$,
and $T$-invariance then forces $(\tau[a])_1=a_1$ or ${\bar a}_1$. Therefore
$M$ is a simple current invariant, and we are done.

(ii):\quad The argument is the same as (i); the exceptional levels here is
$k=(2,16)$ and (2,4).

(iii):\quad Suppose $k_2>16$ (the remaining ten levels $k\in\{4,8,12,16\}
\times\{4,8,12,16\}$ have already been done explicitly). Then as usual looking
at $f^2=(1,3)$ implies $(\tau[a])_2=a_2$ or ${\bar a}_2$, $\forall a\in
\p_L$. Looking at $f^1=(3,1)$ then implies $(\tau[a])_1=a_1$ or ${\bar a}_1$,
for any $a$, except possibly if $k_1=16$. Therefore $M$ will be a
simple current invariant, except when $k_1=16$ and $(\tau[f^1])_1=9$. The
remainder of the argument, namely that in the latter case $M$ must equal
${\cal E}_{16}\otimes {\cal D}_{k_2}$, is as in Thm.6 or Thm.7.

(b) The usual arguments from $(6.1b)$ give us that, assuming $\R_M\not\subset
\I\rho$, $\R^L_M=\R^R_M=\I_L\{(1,1),(11,1)\}$, where either $\I_L=\{0,J^1\}$
or $\I_L=\{0,J^1,J^2,J^{12}\}$. Also, each $M_{\rho b}=M_{b \rho}$ is 0 or 1.

Consider the block diagonal form $M'=\sum |ch_i|^2$. Then Lemma 2 and
Perron-Frobenius tells us that its blocks $B'_i$ will either be a $(2\|
\I_L\|)\times(2\|\I_L\|)$ block of 1's, a $\|\I_L\|\times\|\I_L\|$ block of
2's, or perhaps a $(\|\I_L\|/2)\times(\|\I_L\|/2)$ block of 4's.
By an argument as in (6.10), we get that $M'_{1x,b}=M'_{11x,b}$ etc.
{}From this we find that $M'=
{\cal E}_{28}\otimes{\cal A}_{k_2}$ (if $\|\I_L\|=2$) or $M'={\cal E}_{28}
\otimes {\cal D}_{k_2}$ (if $\|\I_L\|=4$).

Now return to the given $M$. It will correspond to some automorphism $\tau$
of the chiral algebra defined by $M'$. The $S$-matrix of that chiral algebra
can be easily computed, it turns out to be $S'\otimes S''$, where $S'$ is
the $2\times 2$ $S$-matrix defined by ${\cal E}_{28}$, and $S''$ is either
the $S$-matrix for ${\cal A}_{k_2}$ or ${\cal D}_{k_2}$. From these explicit
values one quickly finds that $\tau$ also must factorize in this way (apart
from $k_2=3$), which means $M$ must be in the desired form.

(c) The proof here is similar to that of (b).
\qquad QED \medskip

If both $k_1,k_2$ are odd, then Thm.5 says that the conclusion of Proposition
(a) still holds if we consider there, instead of {\it strongly physical}
invariants $M$, any {\it physical} invariant $M$.

This proposition covers most of the physical invariants of $g^2_k$. It
says that we are done the $g^2_k$ classification if we can show, apart
from some small $k=(k_1,k_2)$ which can be handled individually, that the
following holds:
$$\eqalignno{(a,b)\in\R_M\Rightarrow &\,a\in\{1,k_1'-1\}\,\,\forall k_1,\,\,
{\rm or}&\cr
&\,a\in\{1,3,5\}\,\,{\rm for}\,\,k_1=4,\,\,{\rm or}&\cr
&\,a\in\{1,3,7,9\} \,\,{\rm for \,\,}k_1=8,\,\,{\rm or}&\cr
&\,a\in\{1,5,7,11\}\,\,{\rm for}\,\,k_1=10, \,\,{\rm or}&\cr
&\,a\in\{1,11,19,29\}\,\,{\rm for}\,\,k_1=28.&(7.1)\cr}$$
The reason is that Lemma 3 and $(6.2a$) then constrain $b$ similarly, and
with $T$-invariance $\R_M$ is reduced to the possibilities considered in
the proposition, with three exceptions. They are $k=(10,10)$, (10,28), and
(28,28). The first two are explicitly worked out by computer in \HK{} and \GH,
respectively.

The remaining special case $k=(28,28)$ can be handled by the now-familiar
arguments. If $\R_M\not\subset\{1,11,19,29\}\times\{1,29\}$ and $\R_M
\not\subset\{1,29\}\times\{1,11,19,29\}$, then from $(6.1b$) we get $\R^L_M
=\R^R_M=\{1,11,19,29\}\times\{1,11,19,29\}$, and all $M_{\rho,b}$ and
$M_{a,\rho}$ equal 0 or 1. So $\I_L=\I_R=\{0,J^1,J^2,J^{12}\}$. The (6.10)
argument tells us $M_{1x,1y}=M_{11x,1y}$ etc. Equation $(6.1b)$ tells us
precisely which rows and columns will be non-zero. Each non-zero row and
column of $M$ must sum to 16. Putting all this together, we find that
$M$ is completely fixed once we know whether $M_{17,17}=1$ or 0 -- then
$M={\cal E}_{28}\otimes{\cal E}_{28}$ or $({\cal E}_{28}\otimes
{\cal E}_{28})^c$, resp.

So it suffices to find all levels $k$ where ${\cal K}_M$ does not satisfy
$(7.1$). We can expect there to only be finitely many such $k$ -- they will
be where to find the remaining ${\cal E}_{10}$-like exceptionals.
If we were to follow the approach of \GR, we would have three main tools
to constrain $\R_M$: equations (6.1),(6.2). The analysis is in fact
somewhat simpler here, with one important qualification: there are two
independent levels, $k_1$ and $k_2$, here while in \GR{} we had only one.
This makes the number of different cases to be considered quite large.

There is an alternative, though it requires that we impose the full
machinery of \MS. By analysing the polynomial solutions of the corresponding
Knizhnik-Zamolodchikov equations, Stanev \STA{} has found the list of all
possible chiral extensions of $A_{1,k_1}\oplus A_{1,k_2}$. The only
sporatic levels on this list turn out to be (2,2), (3,1), (6,6), (8,3),
(10,2), (10,10), and (28,8), corresponding to conformal embeddings of
$A_{1,k_1}\oplus A_{1,k_2}$ into $A_3$, $G_2$, $B_4$, $C_3$, $D_4$, $D_5$,
and $F_4$, respectively.

These levels have all been explicitly checked in \HK{} and \GH{}. The
remaining levels will then have $\rho$-couplings given in the proposition
(apart from the 3 exceptionals noted previously). The conclusion is:

\medskip\noindent{{\bf Theorem 8.}} \quad {\it The complete list of strongly
physical invariants of $A_{1,k_1}\oplus A_{1,k_2}$ is given in Sect.2.}

In particular, for each $k=(k_1,k_2)$ there is
$N(k)$ physical invariants, where

\item{(i)} $N(k)=1$ for $k_1=2$ and $k_2$ odd (or vice versa);

\item{(ii)} $N(k)=2$ for $k_1$ odd and $k_2\not\in\{ k_1, 2, 10, 16, 28\}$,
(or vice versa), and $k\ne(3,8)$;

\item{} $N(k)=2$ for $k=(1,1)$;

\item{(iii)} $N(k)=3$ for $k_1$ odd and $k_2\in\{10, 16, 28\}$
(or vice versa), and $k\ne (3,28)$;

\item{} $N(k)=3$ for $k_1=2$ and $k_2\not\in\{ 10,16, 28\}$ is
even (or vice versa);

\item{} $N(k)=3$ for $k=(3,8)$;

\item{(iv)} $N(k)=4$ for $k_1=k_2$ both odd, provided $k\ne (1,1)$;

\item{} $N(k)=4$ for $k=(2,16)$, (2,28) or (3,28);

\item{(v)} $N(k)=6$ for $k_1\ne k_2$ both even, unless either $k_1$ or $k_2$
lies in $\{2, 10, 16, 28\}$;

\item{} $N(k)=6$ for $k=(2,10)$;

\item{(vi)} $N(k)=8$ for $k_1\in\{16,28\}$ and $k_2\not\in\{2,10,16,28\}$
is also even (or vice versa), unless $k=(28,8)$;

\item{(vii)} $N(k)=9$ for $k_1=10$ and $k_2\not\in\{2,10,16,28\}$;

\item{} $N(k)=9$ for $k=(28,8)$;

\item{(viii)} $N(k)=11$ for $k=(16,28)$;

\item{(ix)} $N(k)=12$ for $k_1=k_2$ even, unless $k_1=k_2\in\{2, 4, 6, 8, 10,
16, 28\}$;

\item{} $N(k)=12$ for $k=(10,28)$ or (10,16);

\item{(x)} $N(k)=13$ for $k=(4,4)$, (6,6) or (8,8);

\item{(xi)} $N(k)=22$ for $k=(16,16)$ or (28,28);

\item{(xii)} $N(k)=27$ for $k=(10,10)$.

The term ``vice versa'' in (i), (ii), (iii) and (vi) refers to the obvious
fact that $N(k_1,k_2)=N(k_2,k_1)$.
All the physical invariants for $k_1=k_2$ are explicitly given in \HK;
all those for $k_1\ne k_2$ are in \GH.

\bigskip\bigskip\noindent{{\bf 8. Conclusion}}\bigskip

In this paper we classify all partition functions for $A_{1,k_1}\oplus
A_{1,k_2}$ WZNW theories. The result, given in Sect.2, can be summarized
as follows. There are the analogues of the ${\cal A}$ and ${\cal D}$
series, 2 or 6 of them for each $k=(k_1,k_2)$, depending on whether
or not one of the $k_i$ is odd. If $k_1$ or $k_2$ is 10, 16 or 28, there
are also the invariants that can be built up from the $A_1$ exceptionals
${\cal E}_{10}$, ${\cal E}_{16}$ or ${\cal E}_{28}$, respectively. When
$k_1=k_2$, the conjugations of all these invariants must be included.
Finally, there is one additional exceptional at each level $k=(4,4)$,
(6,6), (8,8), (10,10), (2,10), (3,8), (3,28) and (8,28). The number of
physical invariants for a given $k$ will usually be 2 or 6, but gets
as high as 27 (for $k=(10,10)$).

Our main focus however has not been on $A_1\oplus A_1$, but rather on
the arbitrary rank case $g^r=A_1\oplus\cdots\oplus A_1$, where we have obtained
a number of results (see Thms.1-7). The main remaining obstacle to the
classification for arbitrary rank $r$ is to
find all possible exceptional chiral extensions. To this end, the recent work
of Stanev \STA{} is most interesting, and together with the work in this
paper should permit a classification at least for $g^3$ and $g^4$, along with
``almost every'' level $k$ for each $r>4$. Rank $r=3$
is of particular interest, since it should lead to a classification of all
$su(2)_k\oplus su(2)_\ell/su(2)_{k+\ell}$ GKO cosets \GWA.
In this paper we have found that the most difficult levels for the $g^r_k$
classification seem to be those for which
some $k_i=2$, 4, 8, 10, 12, 16 or 28. But we should be able to handle these
anomolous cases explicitly, as was done in Sect.3 for $k_i=2$, or in the
proposition in Sect.7.

Some of the results of this paper have already found direct application
(see \GH{} and \GWA). More important, much of the techniques developed
in this paper can be carried over to other RCFT classifications -- \eg
Lemma 2 here permits a significant simplification of Sect.5 of \GR. In
fact, thanks to this, it is now possible to rewrite that section of \GR{}
in such a way that that paper finds all {\it physical} invariants of
$su(3)_k$ (previously, for half the levels $k$ some results from \MS{} were
needed, so for those levels
the argument applied to only {\it strongly physical} invariants). In other
words, the classification for $su(3)_k$, $\forall k$, now requires only
(P1)-(P3).

This paper is designed to probe the largely unknown realm of large rank
semi-simple physical invariants. We find enormous numbers of physical
invariants, but most of these are easily tractible, using for example
simple currents. The great hope, albeit one with little hard justification
at present, is that for any affine algebras and levels, irregularities really
only appear when simultaneously the rank {\it and} levels are small; all
other invariants can be obtained from these and from {\it generic} physical
invariants, using the standard constructions. In other words, the hope is that
{\it a finite amount of information captures all physical invariants}.
For example for $su(N)_k$,
the ``generic'' physical invariants would be the ${\cal A}$ and ${\cal D}$
series and their conjugations, together with the conformal embeddings at
$k=N-2$, $N$, and $N+2$;
irregularities have only appeared so far at $(N,k)=(2,10)$, (2,16), (2,28),
(3,9), (3,21), (4,8), (5,5), (6,6), (8,4), (8,10), (9,3) and (16,10).
This paper is consistent with this vision, but much, much work remains.

The $su(2)$ classification yielded the A-D-E problem \CIZ.
The $su(3)$ classification produced the Fermat curve coincidences \RTW.
Both of these hint of a deep, rich structure underlying these problems.
It would be very interesting to discover if any further ``coincidences''
arise, connected to the $su(2)\oplus su(2)$ classification given here.

\bigskip \noindent{{\bf Acknowledgments}}
This work is supported in part by the Natural Sciences and Engineering
Research Council of Canada. During the course of writing this paper,
I have benefitted from conversations with Antoine Coste, Quang Ho-Kim,
Philippe Ruelle, Yassen Stanev and Mark Walton; in particular without Prof.
Ho-Kim's
programming skills it would have been very difficult to complete this
paper. I also appreciate the
hospitality shown
by the IHES and the Carleton mathematics department.

\bigskip\bigskip\noindent{{\bf References}} \bigskip

\item{1.} Bais, F.~A., Bouwknegt, P.~G.: A classification of subgroup
truncations of the bosonic string. Nucl. Phys. {\bf B279} 561-570 (1987);

\item{} Schellekens, A.~N., Warner, N.~P.: Conformal subalgebras of
Kac-Moody algebras. Phys. Rev. {\bf D34} 3092-3096 (1986)

\item{2.} Bernard, D.: String characters from Kac-Moody automorphisms.
Nucl. Phys. {\bf B288} 628-648 (1987);

\item{} Ahn, C., Walton, M.~A.: Spectra on nonsimply-connected group
manifolds. Phys. Lett. {\bf B223} 343-348 (1989)

\item{3.} Bouwknegt, P.: On the construction of modular invariant
partition functions. Nucl. Phys. {\bf B290 [FS20]} 507-526 (1987)

\item{4.} Bouwknegt, P., Nahm, W.: Realizations of the exceptional
modular invariant $A_1^{(1)}$ partition functions. Phys. Lett. {\bf B184}
359-362 (1987)

\item{5.} Cappelli, A.: Modular invariant partition functions of
superconformal theories. Phys. Lett. {\bf B185} 82-88 (1987)

\item{6.} Cappelli, A., Itzykson, C., Zuber, J.-B.: Modular invariant
partition functions in two dimensions. Nucl. Phys. {\bf B280 [FS18]}
445-465 (1987); The A-D-E classification of $A_1^{(1)}$ and minimal
conformal field theories. Commun. Math. Phys. {\bf 113} 1 (1987);

\item{} Kato, A.: Classification of modular invariant partition functions
in two dimensions. Mod. Phys. Lett. {\bf A2} 585 (1987);

\item{} Gepner, D., Qui, Z.: Modular invariant partition functions for
parafermionic theories. Nucl. Phys. {\bf B285} 423-453 (1987)

\item{7.} Cleaver, G.~B., Lewellen, D.~C.: On Modular invariant
partition functions for tensor products of conformal field theories. Phys.
Lett. {\bf B300} 354-360 (1993)

\item{8.} Coste, A., Gannon, T.: Galois symmetry in RCFT. Phys. Lett. {\bf B}
(to appear)

\item{9.} Degiovanni, P.: Modular invariance with a non simple symmetry
algebra. Nucl. Phys. {\bf B} (Proc. Suppl.) {\bf 5B} 71-86 (1988)

\item{10.} Fuchs, J., Klemm, A., Schmidt, M.~G., Verstegen, D.: New
exceptional $(A_1^{(1)})^{\oplus_r}$ invariants and the associated Gepner
models. Int. J. Mod. Phys. {\bf A7} 2245-2264 (1992)

\item{11.} Gannon, T.: WZW commutants, lattices, and level-one partition
functions. Nucl. Phys. {\bf B396} 708-736 (1993)

\item{12.} Gannon, T.: Partition functions for heterotic WZW
 conformal field theories. Nucl. Phys. {\bf B402} 729-753 (1993)

\item{13.} Gannon, T.: The classification of affine SU(3)
modular invariant partition functions. Commun. Math. Phys. (to appear)

\item{14.} Gannon, T., Ho-Kim, Q.: The low level modular invariant
partition functions of rank-two algebras. Int. J. Mod. Phys. {\bf A} (to
appear)

\item{15.} Gannon, T., Ho-Kim, Q.: The rank-four heterotic
modular invariant partition functions. Preprint, IHES

\item{16.} Gannon, T., Walton, M.~A.: The classification of
GKO modular invariants (work in progress)

\item{17.} Gantmacher, F.R.: The theory of matrices Vol. II. New York:
Chelsea Publishing Co. 1964

\item{18.} Gato-Rivera, B., Schellekens, A.~N.: Complete classification of
simple current automorphisms. Nucl. Phys. {\bf B353} 519-537 (1991);

\item{} Schellekens, A.~N.: Fusion rule automorphisms from integer spin
simple currents. Phys. Lett. {\bf B244} 255-260 (1990)

\item{19.} Gato-Rivera, B., Schellekens, A.~N.: Complete classification
of simple current modular invariants for RCFT's with a center $(\Z_p)^k$.
Commun. Math. Phys. {\bf 145} 85-121 (1992);

\item{} Kreuzer, M., Schellekens, A.~N.: Simple currents versus orbifolds
with discrete torsion -- a complete classification. Nucl. Phys. {\bf 411}
97-121 (1994)

\item{20.} Gepner, D., Witten, E.: String theory
on group manifolds. Nucl. Phys. {\bf B278} 493-549 (1986)

\item{21.} Itzykson, C.: Level one Kac-Moody characters and modular
invariance. Nucl. Phys. (Proc. Suppl.) {\bf 5B} 150-165 (1988);

\item{} Degiovanni, P.: Z/NZ conformal field theories.
Commun. Math. Phys. {\bf 127} 71-99 (1990)

\item{22.} Ka{\v c}, V.~G.: Infinite Dimensional Lie Algebras, 3rd ed.
Cambridge: Cambridge University Press 1990

\item{23.} Ka{\v c}, V.~G., Wakimoto, M.: Modular and conformal invariance
constraints in representation theory of affine algebras. Adv. Math. {\bf 70}
156-236 (1988)

\item{} Verstegen, D.: Conformal embeddings, rank-level duality, and
exceptional modular invariants. Commun. Math. Phys. {\bf 137} 567-586 (1991)

\item{} Walton, M.~A.: Conformal branching rules and modular invariants.
Nucl. Phys. {\bf B322} 775-790 (1989)

\item{24.} Moore, G., Seiberg, N.: Naturality in conformal field theory.
 Nucl. Phys. {\bf B313} 16-40 (1989)

\item{25.} Ruelle, Ph., Thiran, E., Weyers, J.: Implications
of an arithmetical symmetry of the commutant for modular invariants. Nucl.
Phys. {\bf B402} 693-708 (1993)

\item{26.} Schellekens, A.~N., Yankielowicz, S.: Extended chiral algebras and
modular invariant partition functions. Nucl. Phys. {\bf B327} 673-703 (1989)

\item{27.} Stanev, Y.: Local extensions of the $SU(2)\times SU(2)$ conformal
current algebra. (in preparation); (private communication)

\item{28.} Verstegen, D.: New exceptional modular invariant partition
functions for simple Kac-Moody algebras. Nucl. Phys. {\bf B346} 349-386 (1990)

\item{29.} Witten, E.: Non-abelian bosonization in two dimensions.
 Commun. Math. Phys. {\bf 92} 455-472 (1984);

\item{} Novikov, S.~P.: Usp. Mat. Nauk {\bf 37} 3 (1982)

\end